\def\arxivversion{}     

\ifdefined\arxivversion
    \documentclass[sigconf]{acmart}
\else
    \documentclass[sigconf,review,anonymous]{acmart}
\fi

\usepackage{tabu} 
\usepackage{xcolor}
\usepackage{caption}
\usepackage{subcaption}
\usepackage{import}
\usepackage{multirow}
\usepackage{graphicx}
\usepackage{cleveref}
\usepackage{soul}
\usepackage{xcolor,colortbl}
\usepackage{pifont}
\usepackage{arydshln}
\usepackage{changepage}
\usepackage{mdframed}

\usepackage{tikz}
\definecolor{customRed}{HTML}{C00000}

\newcommand{\system}{SceneScout}

\ifdefined\arxivversion
    \newcommand{\maps}{Apple Maps APIs~\cite{apple_maps_server_api}}
\else
    \newcommand{\maps}{Maps APIs\footnote{We omitted the identity of the specific map services we used to maintain anonymity.}}
\fi

\newcommand{\user}{Maya}

\newcommand{\circlelabel}[2]{%
    {\small{#2}}%
}

\AtBeginDocument{%
  \providecommand\BibTeX{{%
    \normalfont B\kern-0.5em{\scshape i\kern-0.25em b}\kern-0.8em\TeX}}}

\ifdefined\arxivversion
    \setcopyright{rightsretained}
    \acmConference[arXiv]{2025}{April}{2025}
    \acmDOI{xx.xxxx/xxxxxxx.xxxxxxx}
    \acmISBN{xxx-x-xxxx-xxxx-x/xx/xx}
\else
    \copyrightyear{2025}
    \acmYear{2025}
    \setcopyright{acmlicensed}
    \acmConference[UIST '25]{The 38th Annual ACM Symposium on User Interface Software and Technology}{September 28--October 1, 2025}{Busan, Korea}
    \acmBooktitle{The 38th Annual ACM Symposium on User Interface Software and Technology (UIST '25), September 28--October 1, 2025, Busan, Korea}
    \acmDOI{10.1145/1234567.12345678}
    \acmISBN{123-4-5678-1234-5/25/09}
\fi

\begin{document}




\title[\system: Towards AI Agent-driven Access to Street View Imagery for Blind Users]{\system: Towards AI Agent-driven Access \\to Street View Imagery for Blind Users}



%



\author{Gaurav Jain}
\authornote{Work done during internship at Apple.}
\affiliation{
  \institution{Columbia University}
  \city{New York}
  \state{NY}
  \country{USA}
}
\email{gaurav@cs.columbia.edu}

\author{Leah Findlater}
\affiliation{
  \institution{Apple}
  \city{Seattle}
  \state{WA}
  \country{USA}
}
\email{lfindlater@apple.com}

\author{Cole Gleason}
\affiliation{
  \institution{Apple}
  \city{Seattle}
  \state{WA}
  \country{USA}
}
\email{cole_gleason@apple.com}



\begin{abstract}
People who are blind or have low vision (BLV) may hesitate to travel independently in unfamiliar environments due to uncertainty about the physical landscape. While most tools focus on in-situ navigation, those exploring pre-travel assistance typically provide only landmarks and turn-by-turn instructions, lacking detailed visual context. Street view imagery, which contains rich visual information and has the potential to reveal numerous environmental details, remains inaccessible to BLV people. In this work, we introduce \textit{\system}, a multimodal large language model (MLLM)-driven AI agent that enables accessible interactions with street view imagery. \system\ supports two modes: (1) Route Preview, enabling users to familiarize themselves with visual details along a route, and (2) Virtual Exploration, enabling free movement within street view imagery. Our user study ($N=10$) demonstrates that \system\ helps BLV users uncover visual information otherwise unavailable through existing means. A technical evaluation shows that most descriptions are accurate (72\%) and describe stable visual elements (95\%) even in older imagery, though occasional subtle and plausible errors make them difficult to verify without sight. We discuss future opportunities and challenges of using street view imagery to enhance navigation experiences.
\end{abstract}



\begin{teaserfigure}
    \centering
    \includegraphics[width=0.99\linewidth]{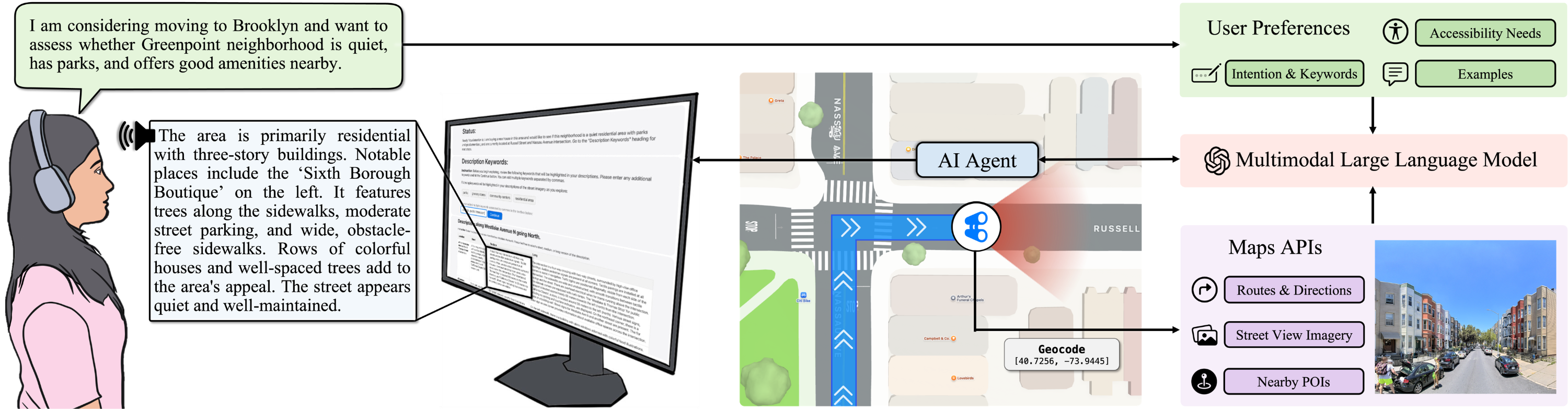}
    \caption{The \system\ prototype system. A blind user accessing street view imagery through textual descriptions provided by \system\ to assess a new neighborhood for potential relocation. \system\ generates these descriptions by virtually grounding an AI agent within the physical environment using maps APIs, enabling the agent to perceive real-world surroundings. This perceived visual and map information, along with BLV users’ preferences (\emph{e.g.} intention and accessibility needs), is processed by a multimodal large language model to extract meaningful information, which blind users access through textual descriptions displayed on \system’s web interface.}
    \Description{SceneScout system overview, illustrating the interaction between five main components: (1) the blind user, (2) a web interface, (3) user preferences (e.g., accessibility needs, intentions, keywords), (4) an AI agent navigating street view imagery, and (5) a multimodal large language model. The flow begins with the user providing a spoken query, which is used to define user preferences. These preferences, along with data from the AI agent--who explores neighborhoods using Maps APIs (routes, street imagery, POIs) and geolocation--are fed into the language model. The model synthesizes this multimodal input into a grounded, personalized description, which is delivered back to the user via the web interface. Arrows in the diagram indicate the directional flow of data between these components.}
  \label{fig:teaser}
\end{teaserfigure}


\maketitle

\section{Introduction}
\label{sec:introduction}
Navigating outdoors is a major challenge for blind and low vision (BLV) people\footnote{Disability language is nuanced and shaped by regional, cultural, and individual preferences~\cite{sharif_should_2022}. In this work, we use the term \emph{blind and low vision (BLV) people} to refer to individuals who rely on screen readers to access visual content on computers.}. Existing solutions assist BLV users with various aspects of navigation, including guiding them to destinations~\cite{ahmetovic_navcog_2016, sato_navcog3_2019, jain_streetnav_2024, jain_towards_streetnav_2023}, alerting them of nearby obstacles~\cite{guerreiro_cabot_2019, kuribayashi_corridor_2022, jain_streetnav_2024}, highlighting points of interest (POIs)~\cite{blindsquare, kacorri_supporting_2016, sato_navcog3_2019, microsoft2018soundscape}, and facilitating street crossing~\cite{oko, mascetti_zebrarecognizer_2016, ahmetovic_zebra_2015, jain_streetnav_2024}. Despite advancements in in-situ navigation assistance tools, BLV users may still hesitate to navigate unfamiliar environments independently. This hesitation stems in part from uncertainty about the built environment’s physical characteristics (\emph{e.g.} sidewalks, curb ramps, crosswalks), which current tools do not fully address~\cite{helal_blind_2008, williams_pray_2013, williams_just_2014, jain_i_2023, cushley_unseen_2023}.

To complement in-situ navigation assistance tools, research has explored approaches for helping users acquire spatial and route knowledge before navigation~\cite{montello_spatial_2001, brock_interactive_2013, guerreiro_virtual_2020, guerreiro_virtual_2017}. Approaches such as route previews~\cite{guerreiro_virtual_2020} enable users to form mental maps of their routes before physically navigating the environment. However, they often focus on navigation instructions and POIs along the route, while overlooking critical details about the accessibility of the built environment. Features such as tactile paving, curb ramps, and audible pedestrian signals (APSs) significantly influence BLV users’ willingness to navigate unfamiliar areas~\cite{saha_project_2019, kameswaran_understanding_2020, jain_i_2023}, yet this information is rarely included in existing solutions. Thus, there is a need to explore approaches that provide BLV users with a holistic understanding of the environment beforehand, ultimately improving their confidence during navigation~\cite{kitchin_techniques_1997, Papadopoulos2017Cognitive}.

Street view imagery, a rich source of information about the outdoor environment, offers a potential solution. It consists of panoramic photographs captured at street level, providing a visual representation of streets, sidewalks, and surrounding infrastructure. Sighted people frequently use tools like Apple Maps~\cite{apple_maps} and Google Maps~\cite{google_maps} to \textit{visually inspect} street view imagery, allowing them to assess physical surroundings and contextualize the navigation instructions by themselves. However, this resource remains largely inaccessible to BLV users. Although prior research has shown the utility of street view imagery for identifying accessibility issues~\cite{saha_project_2019, hara_tohme_2014, hara_improving_2015}, there is still limited work to make street view imagery directly usable by BLV users.

Many questions arise before facilitating direct access to street view imagery for BLV users: What various types of information could be useful to BLV people from street view imagery? How do their preferences for information from street view imagery change by scenario? How might we design a system that can perceive and move within street view imagery to enable accessible interactions? To what extent can such a system address BLV people's needs?

In this work, we address these questions by developing \textit{\system}, a prototype system that uses a multimodal large language model (MLLM)-driven AI agent to enable blind-accessible interactions with street view imagery. Based on a literature review and feedback from BLV colleagues, we designed \system\ to support two primary interaction modes: (1) \textit{Route Preview}, which familiarizes users with visual details along a specific route, and (2) \textit{Virtual Exploration}, which enables free movement within street view imagery. Figure~\ref{fig:teaser} illustrates how \system\ grounds an AI agent within the physical world using maps APIs, allowing it to perceive the environment, reason about BLV users’ preferences, and generate personalized textual descriptions based on the users' intentions. These descriptions are presented via \system’s web interface.

We conducted a user study with $N=10$ BLV participants using a mixed-methods, scenario-based design~\cite{rosson2002scenario}. Participants first interacted with \system’s two interaction modes and then answered questions in a semi-structured interview. Findings show that \system's AI agent effectively surfaces relevant information from street view imagery, enabling BLV users to uncover visual details about the built environment that would otherwise remain inaccessible. Participants expressed a strong interest in integrating street view imagery into their navigation workflows, both pre-travel and in situ. However, certain aspects of \system\ did not meet participants' expectations. For example, the descriptions lacked spatial precision, occasionally made assumptions about users’ abilities and the environment, and did not offer the level of personalization desired. We discuss future opportunities for achieving personalization at scale in Section~\ref{sec:discussion}.

To assess \system’s performance, we conducted a technical evaluation to quantify the dimensions highlighted in the user study. Specifically, we evaluated the generated descriptions based on information type, correctness, error type, consistency, redundancy, and relevance. We found that while most descriptions were accurate (72\%), the errors were often plausible additions, textual errors, or spatial mix-ups (16\%) that may be undetected by BLV users who cannot see the images. The descriptions did describe elements that were likely or mostly likely (95\%) to remain consistent over time, meaning the street imagery can remain useful even if not updated on a very frequent basis. While the performance of models will certainly improve, this evaluation indicates we should take care in the design of responses to convey relevant information from street view images to BLV users.

In summary, this work contributes:\begin{itemize}%
    \item \system, a prototype that explores the use of MLLM-driven AI agents to make street view imagery accessible for BLV users. Its two interaction modes, Route Preview and Virtual Exploration, support users' pre-travel planning.
    \item A user study investigating how BLV people use \system\ to access street view imagery. We find that \system\ helps users uncover relevant visual details, but limitations in spatial precision and unsupported assumptions about the environment led to hesitation in fully trusting it.
    \item A technical evaluation assessing the accuracy and reliability of MLLM-driven AI agents in surfacing relevant and accessible information to BLV users. We find that the conversational nature of these models can lead to some subjectivity in descriptions. Although most descriptions are correct, the errors that are present tend to be plausible and hard to discern without the ability to see the images.
\end{itemize}%

\section{Related Work}
\label{sec:related_work}


\subsection{In Situ Navigation Assistance}
Most navigation tools for BLV people focus on in situ real-time guidance, providing turn-by-turn directions~\cite{ahmetovic_navcog_2016, sato_navcog3_2019, jain_streetnav_2024, jain_towards_streetnav_2023}, alerting users of nearby obstacles and POIs~\cite{guerreiro_cabot_2019, kuribayashi_corridor_2022, jain_streetnav_2024, blindsquare, kacorri_supporting_2016, sato_navcog3_2019, microsoft2018soundscape}, and helping users cross streets~\cite{oko, mascetti_zebrarecognizer_2016, ahmetovic_zebra_2015, jain_streetnav_2024}. Although helpful, these systems primarily provide localized navigation cues rather than broader spatial awareness. For example, StreetNav~\cite{jain_streetnav_2024} helps users detect and bypass obstacles, but only within their immediate vicinity, leaving the overall structure of the environment unclear. This limitation highlights the need to provide BLV users with a comprehensive spatial understanding before entering unfamiliar areas~\cite{helal_blind_2008, williams_pray_2013}. Our work addresses this gap by enabling BLV users to virtually explore environments via street view imagery, offering a holistic view of what the environment looks like to proactively build stronger mental models and ultimately navigate confidently.

\subsection{Pre-travel Navigation Assistance}
BLV people often plan routes in advance to anticipate potential challenges and navigate with confidence in unfamiliar environments~\cite{williams_pray_2013, montello2001spatial, brock2013interactive, denis2017space, kitchin_techniques_1997, Papadopoulos2017Cognitive}. Mainstream navigation apps like Google Maps~\cite{google_maps} and Apple Maps~\cite{apple_maps} can help by offering textual, step-by-step directions (\emph{e.g.} \textit{`Turn left onto Park Avenue'}), but these route-focused previews provide limited details about the environment~\cite{kameswaran_understanding_2020}. To offer richer spatial information, researchers have developed tactile maps and 3D-printed models that provide detailed information via touch~\cite{herman1983constructing, wiener2010foundations, giraud2017map, leo2017effect, taylor2016customizable}. Researchers have also proposed virtual navigation systems that convey spatial layouts through audio-haptic feedback. Some of these systems simulate real-world movement using virtual canes or walking-in-place techniques but often require additional hardware~\cite{evett_accessible_2008, Maidenbaum2013Increasing, Kreimeier2019first, zhao_enabling_2018}. Others leverage touchscreen displays with audio-haptic feedback, enabling virtual previews on widely available smartphones~\cite{Guerreiro2015Blind, kane2011access, su2010timbremap, Brock2015Interactivity, Ducasse2018, evett_accessible_2008, Lahav2018virtual, LAHAV2008haptic, sanchez2010audio, PICINALI2014393, guerreiro_virtual_2017, guerreiro_virtual_2020}. 
However, many existing solutions have limited resolution or emphasize basic route information and POIs rather than detailed spatial context. Our work builds on these approaches by investigating how street view imagery, a high-fidelity visual representation of the actual physical environment, can enhance BLV users' pre-travel navigation planning.

\subsection{Street View Imagery for Accessibility}
Street view imagery has been widely explored as a proxy for the built environment, particularly in identifying accessibility issues. 
Project Sidewalk~\cite{saha_project_2019, saha_designing_2022, saha_visualizing_2022, Hara_Scalable_2016}, for example, introduced the concept of crowdsourcing accessibility information like curb ramps and sidewalk issues from Google Street View images~\cite{google_street_view}. Some approaches have also combined machine learning with crowdsourcing to partially automate the annotation workflow~\cite{li_labelaid_2024, hara_tohme_2014, duan_scaling_2022, hosseini_towards_2022, weld_deep_2019}. Collectively, prior work has demonstrated the feasibility of using street view imagery as a data source for assessing the environment's accessibility.
However, current approaches typically generate curated outputs for applications such as accessibility maps or urban planning resources~\cite{Li_Accessibility_2025}. BLV people generally do not directly interact with street view imagery, relying instead on preprocessed information like accessible bus stop locations~\cite{hara_improving_2015} or pedestrian route recommendations~\cite{loitsch_accessiblemaps_2020}.
Our work differs by providing BLV users with direct access to street view imagery, empowering them to obtain meaningful information directly from the imagery itself. 
This approach moves away from access to predefined data from street view imagery towards an interactive experience with it.


\subsection{Multimodal AI Agents for Navigation}
Recent advances in vision-language models have led to the development of AI agents capable of real-world navigation. Benchmarks such as Touchdown~\cite{chen_touchdown_2019} use Google Street View panoramas to assess how well these agents follow natural language instructions and visually locate destinations in a complex urban environment. Researchers have explored various methods to build agents that can independently navigate such settings. Schumann et al.~\cite{schumann_velma_2024} proposed an embodied agent that verbalizes its visual observations at each step, thereby providing a dynamic natural language context for its navigation policy. Balata et al.~\cite{balata_automatically_2016} focused on spatially grounded instructions that automatically enrich route directions with salient landmarks and spatial cues to better align with the environment. Complementing these language-centric strategies, Yang et al.~\cite{yang_v-irl_2024} leverage a simulated environment with street view imagery, effectively bridging the gap between virtual training and real-world complexity. Collectively, these advances demonstrate the feasibility of developing multimodal AI agents with enhanced vision-language reasoning for real-world tasks. Our work builds on these advancements by taking a human-centered approach, adapting multimodal AI agents to support BLV users. Rather than focusing on the computational understanding of virtual environments, we design interactions that enable BLV users to access and interpret street view imagery independently.

\section{The \system\ System}
\label{sec:system}

\begin{figure*}[t]
    \centering
    \includegraphics[width=0.99\linewidth]{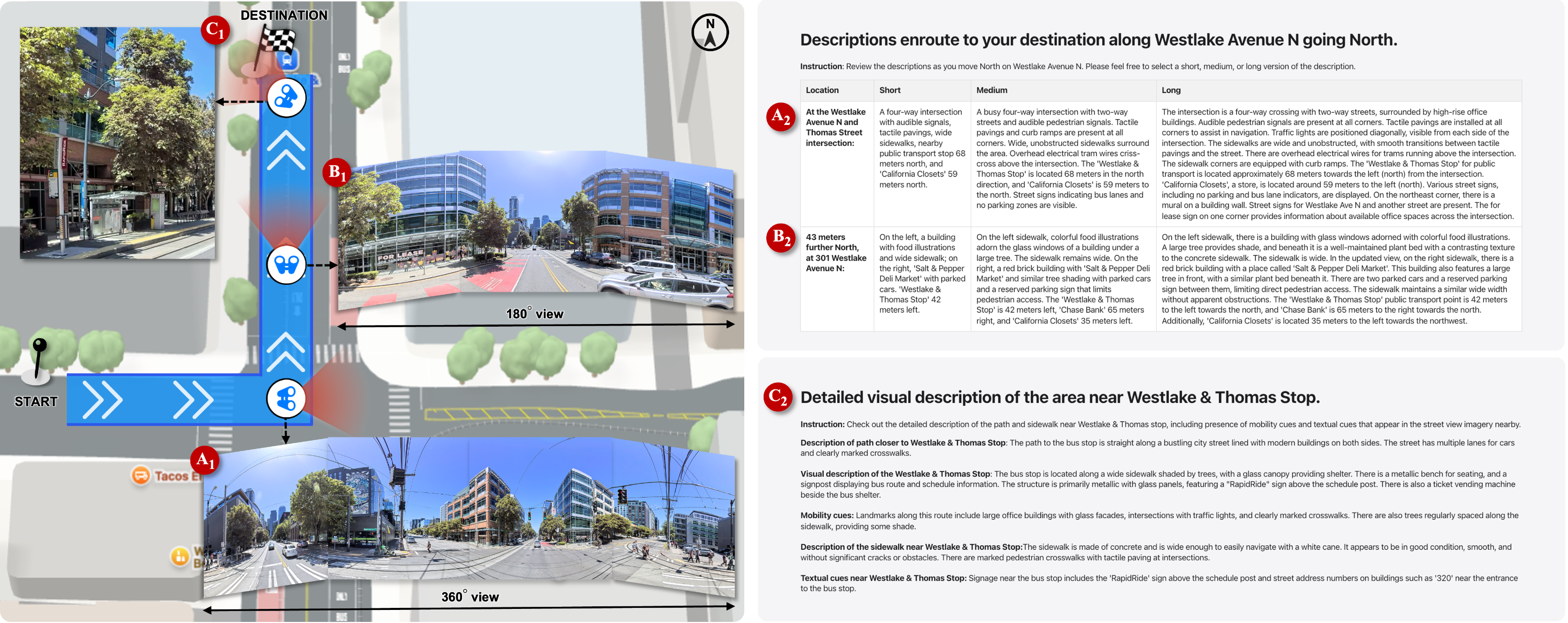}
    \caption{The \textit{Route Preview} interaction mode in \system, enabling BLV users to familiarize themselves with a route before traveling. On the left, an illustration shows the AI agent navigating the street view imagery along the route to a nearby bus stop, while on the right is \system’s web interface. BLV users set a start and destination, which triggers the agent to automatically retrieve relevant street view imagery (A$_1$–B$_1$) and generate step-by-step descriptions (A$_2$–B$_2$) along the route. Finally, it provides a detailed visual description of the destination (C$_2$) based on the destination's street view images (C$_1$), assisting users with last-few-meters wayfinding. Appendix~\ref{sec:route_preview_ui_text} includes the web interface’s text in an accessible format.}
    \Description{SceneScout's Route Preview interaction mode. On the left, a top-down map displays a sidewalk route from a start point to a bus stop destination, with labeled positions A₁, B₁, and C₁ marking points along the path. At each point, panoramic street view images are shown: A₁ (starting point), B₁ (midpoint), and C₁ (destination). The AI agent icon moves along this path, triggering the retrieval of relevant imagery. On the right, the SceneScout web interface displays corresponding step-by-step textual descriptions: A₂, B₂ (landmarks and route features), and a detailed destination description at C₂. These descriptions vary in length (short, medium, long) and highlight mobility cues, sidewalk quality, landmarks, and tactile signals. Arrows and layout emphasize the directional flow from the user-defined route to imagery acquisition, description generation, and user interface output.}\label{fig:route_preview_ui}
\end{figure*}

\system\ is a prototype system that leverages a MLLM-driven AI agent to support accessible interactions with street view imagery for BLV users. \system\ offers two primary modes of interaction through its web interface, (1) \textit{Route Preview}: helping users gain familiarity with the visual details along routes to a destination, and (2) \textit{Virtual Exploration}: enabling users to freely move within street view imagery. 
In the following sections, we discuss the design goals for \system\ (Section~\ref{sec:design_rationale}), the two interaction modes it enables (Section~\ref{sec:interaction_modes}), and the implementation details (Section~\ref{sec:implementation_details}).


\subsection{Design Goals}
\label{sec:design_rationale}
\system's interaction modes embody two design goals identified through a literature review on navigation assistance systems. To ensure we follow a human-centered design process, a critical part of accessibility design~\cite{katta_2020_nothing}, we further refined our design goals through input and feedback from BLV colleagues. We formulate the design goals as follows:

\begin{itemize}
    \item[\textbf{DG1.}] \textbf{Support gaining familiarity with routes and destinations.} 
    Prior work highlights the importance of pre-travel planning for BLV users, showing that access to detailed environmental information aids mental map formation~\cite{nadel_hippocampus_1980, tversky_spatial_1991, ggolledge_cognitive_2000, ottink_cognitive_2022}, helps anticipate challenges~\cite{kameswaran_understanding_2020, jain_i_2023, williams_pray_2013, williams_just_2014, banovic_uncovering_2013}, and helps users feel more confident~\cite{guerreiro_virtual_2017, guerreiro_virtual_2020, abd_hamid_facilitating_2013, clemenson_rethinking_2021}. While broad spatial and accessibility cues are useful for general navigation, last-few-meters wayfinding~\cite{saha_closing_2019} requires finer-grained details such as entrances, storefronts, and nearby landmarks~\cite{gleason_footnotes_2018, jain_streetnav_2024, jain_towards_streetnav_2023}. BLV users should be able to access this layered information through street view imagery to gain familiarity with the routes and  prepare for both navigation along the route and precise arrival at the destination. 
    \vspace{0.1cm}
    
    \item[\textbf{DG2.}] \textbf{Promote exploratory freedom within street view imagery.} While most tools focus on route-based navigation~\cite{ahmetovic_navcog_2016, sato_navcog3_2019, guerreiro_cabot_2019}, recent work highlights exploration as a critical yet under-supported aspect of BLV navigation~\cite{jain_i_2023}. To support exploration with street view imagery, users should be able to freely move in any direction and tailor information to their goals and situational context. This flexibility enables more spontaneous navigation, such as discovering new amenities along less-traveled routes or in unfamiliar locations~\cite{jain_i_2023}.
\end{itemize}

\subsection{Interaction Modes}
\label{sec:interaction_modes}
To illustrate \system's two interaction modes, we follow a persona of a fictional user, \user. She is an engineer based in a large US city who is congenitally blind and relies on screen readers to access computers. To navigate outdoors, \user\ uses a white cane alongside assistive technology applications. We present example usage scenarios for \system's two interaction modes, Route Preview (addressing \textbf{DG1}) and Virtual Exploration (addressing \textbf{DG2}), from \user's perspective.


\subsubsection{\textbf{Route Preview.}}
\label{sec:route_preview_interaction_mode}
\user\ plans to take the bus after work to visit a friend. While she typically walks to a familiar stop to commute home, the bus to her friend’s house departs from a stop she has not used before. Uncertain about the accessibility of the route and hesitant to navigate an unfamiliar path, \user\ turns to \system\ to preview the journey. Figure~\ref{fig:route_preview_ui} illustrates how the route preview interaction mode supports \user\ in this scenario. Appendix~\ref{sec:virtual_exploration_ui_text} includes the figure's text in an accessible format.

\textit{Descriptions along the route.}
\user\ sets her office as the starting point and the bus stop as the destination, a three-minute walk away. \system\ analyzes the route using street view imagery and map metadata (Figure~\ref{fig:route_preview_ui} \textit{left}), and presents the output through a screen reader-accessible web interface (Figure~\ref{fig:route_preview_ui} \textit{right}). Descriptions are organized in a table where each row corresponds to a $30$ to $40$ meter segment. This interval was selected to minimize visual overlap between consecutive panoramas. For each segment, \system\ provides short, medium, and long descriptions based on nearby POIs and contextual details. Prior work highlights the value of customizable levels of detail in accessible image and video content~\cite{huh_genassist_2023, van_daele_making_2024}. This format supports quick skimming, with the option to explore specific segments in more detail as needed.

To generate these descriptions, \system\ adaptively processes different fields of view from street view panoramas. At intersections, it uses a complete $360^\circ$ view to provide a comprehensive understanding of traffic flow and controls, which are crucial for safe crossing. Figure~\ref{fig:route_preview_ui}\circlelabel{customRed}{A${}_1$} shows this view, used to generate the description in Figure~\ref{fig:route_preview_ui}\circlelabel{customRed}{A${}_2$}. Along other segments, it processes a $180^\circ$ forward-facing view, simulating a pedestrian’s perspective. Figure~\ref{fig:route_preview_ui}\circlelabel{customRed}{B${}_1$} shows this view, used for the description in Figure~\ref{fig:route_preview_ui}\circlelabel{customRed}{B${}_2$}.

\begin{figure*}[t]
    \centering
    \includegraphics[width=0.99\linewidth]{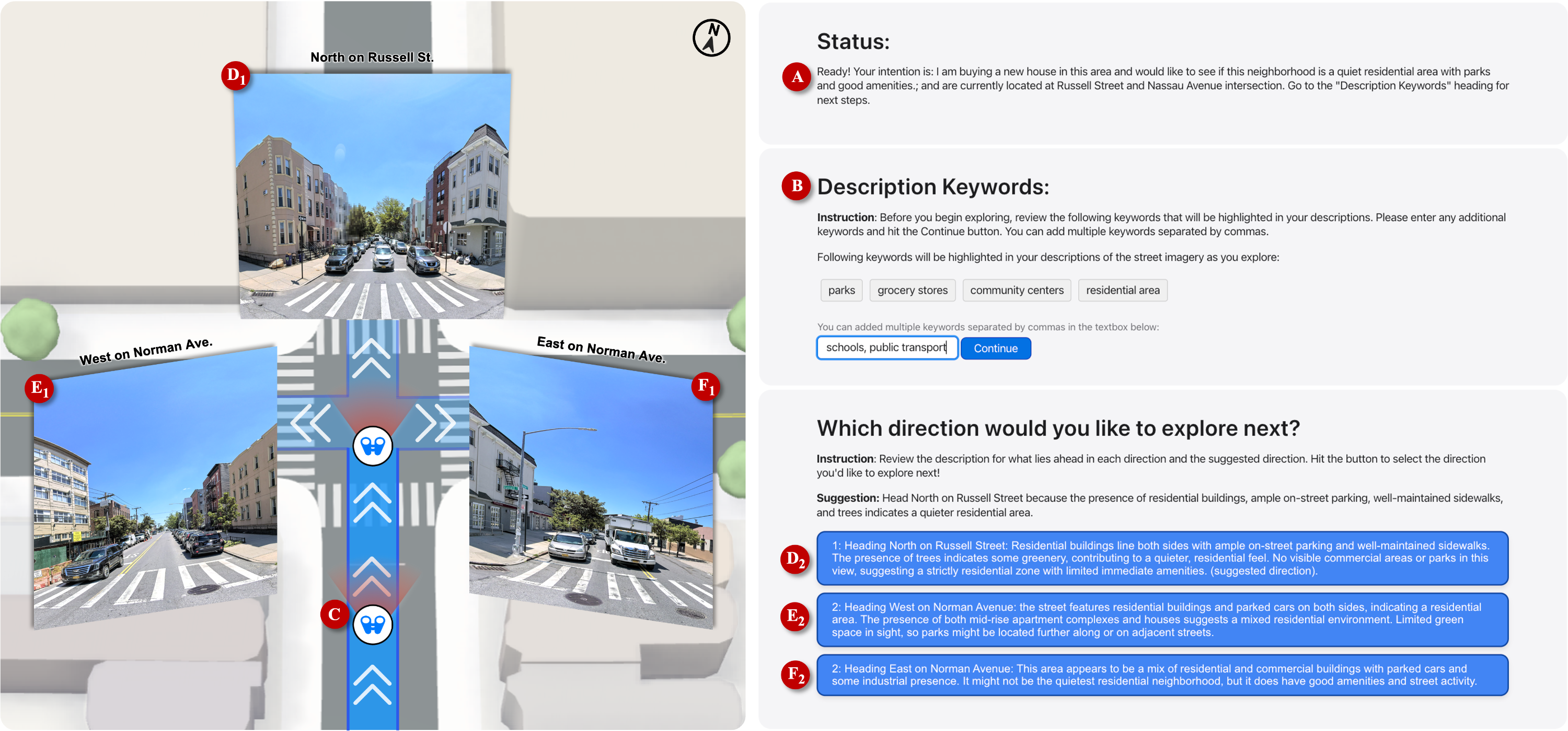}
    \caption{The \textit{Virtual Exploration} interaction mode in \system, enabling BLV users to freely explore street view. On the left, an illustration shows the AI agent's (C) movement within street view, while on the right is \system's web interface. BLV users first specify their intent (A) and relevant keywords (B), which guide the descriptions generated from street view imagery. At intersections, users receive descriptions (D$_2$--F$_2$) of each possible direction (D$_1$--F$_1$) and select where to explore next. The agent then moves accordingly, generating step-by-step descriptions tailored to the user’s intent, creating an interactive and personalized exploration experience. Appendix~\ref{sec:virtual_exploration_ui_text} includes the web interface's text in an accessible format.}
    \Description{SceneScout’s Virtual Exploration interaction mode. On the left, a top-down intersection map shows the AI agent’s position (C) at the center, with three directional options: North on Russell Street (D₁), West on Norman Avenue (E₁), and East on Norman Avenue (F₁). Each direction is accompanied by a corresponding street view image and description. On the right, the SceneScout web interface is divided into three sections: (A) user intent (e.g., searching for a quiet residential area with parks), (B) selected keywords (e.g., “parks,” “grocery stores,” “residential area”), and (D₂–F₂) suggested next directions with descriptive summaries based on the street view imagery. Arrows and spatial layout highlight how the user-defined intent and keywords guide the AI agent’s exploration, enabling SceneScout to generate personalized, step-by-step descriptions of each possible direction for further navigation.}
    \label{fig:virtual_exploration_ui}
\end{figure*}

\textit{Detailed description near the destination.}
After reviewing the route, \user\ receives a detailed description of the environment near the bus stop. This is organized into key categories: (i) the path leading to the stop, (ii) visual features such as a glass shelter with seating, (iii) suggested mobility cues like nearby trees or bins, (iv) sidewalk characteristics, and (v) textual signage, for example, a sign labeled ``RapidRide.'' These categories were informed by prior work~\cite{saha_closing_2019} and feedback from BLV colleagues. This structure helps \user\ identify essential visual and spatial cues for precise navigation. Figure~\ref{fig:route_preview_ui}\circlelabel{customRed}{C${}_1$} shows the street view image used to generate the destination-focused description in Figure~\ref{fig:route_preview_ui}\circlelabel{customRed}{C${}_2$}.

\subsubsection{\textbf{Virtual Exploration.}}
\label{sec:virtual_exploration_interaction_mode}

\user, preparing to relocate to New York City for a new job, is considering moving to the Greenpoint neighborhood in Brooklyn. To evaluate whether this area is worth visiting and possibly living in, \user\ wants to explore the neighborhood. \user\ turns to \system\ and uses it to virtually explore the neighborhood as part of her relocation planning. Figure~\ref{fig:virtual_exploration_ui} illustrates how the virtual exploration interaction mode supports \user\ in this scenario. Appendix~\ref{sec:virtual_exploration_ui_text} includes the text from Figure~\ref{fig:virtual_exploration_ui}'s web interface in an accessible format.


\textit{Defining exploration intent and keywords.}
To begin, \user\ specifies her intention for exploration in natural language (Figure~\ref{fig:virtual_exploration_ui}\circlelabel{customRed}{A${}_{}$}) and provides keywords for \system\ to emphasize in its descriptions of the street view imagery. \user\ enters: ``\textit{I am buying a new house in this area and would like to see if this neighborhood is a quiet residential area with parks and good amenities.}'' To further clarify her priorities, \user\ can provide a set of keywords to guide the descriptions. \system\ includes an initial set of keywords to which \user\ adds: schools and public transport, as shown in Figure~\ref{fig:virtual_exploration_ui}\circlelabel{customRed}{B${}_{}$}. \system\ then processes the street view imagery to generate descriptions of a block (Figure~\ref{fig:virtual_exploration_ui}\circlelabel{customRed}{C${}_{}$}), offering three verbosity levels similar to the route previews discussed in Section~\ref{sec:route_preview_interaction_mode} but tailored to \user's intention and keywords. 

\textit{Choosing directions for further exploration.}
As \user\ explores the neighborhood virtually, she encounters intersections where decisions about the next direction to explore must be made. At these points, \system\ first fetches the street view imagery for each possible direction, as shown in Figure~\ref{fig:virtual_exploration_ui}\circlelabel{customRed}{D${}_1$}--\circlelabel{customRed}{F${}_1$}. Then, it provides a summary of what lies in each direction, emphasizing features that align with \user's specified intention and keywords as shown in Figure~\ref{fig:virtual_exploration_ui}\circlelabel{customRed}{D${}_2$}--\circlelabel{customRed}{F${}_2$}. For example, since ``\textit{quiet streets}'' is a priority for \user, \system\ might highlight directions with less traffic or appear more residential.

To assist in decision-making, \system\ suggests a direction that best matches \user's goals. However, \user\ retains full agency and can override the suggestion to choose an alternative direction. By guiding \user\ through iterative decision points at intersections, \system\ fosters a dynamic exploration process that provides BLV users with an experience similar to how sighted users might navigate through street view imagery. This cycle repeats as \system\ continues to generate block-by-block descriptions, prompting \user\ to choose a direction at each intersection. 

\begin{figure*}[t]
    \centering
    \includegraphics[width=\linewidth]{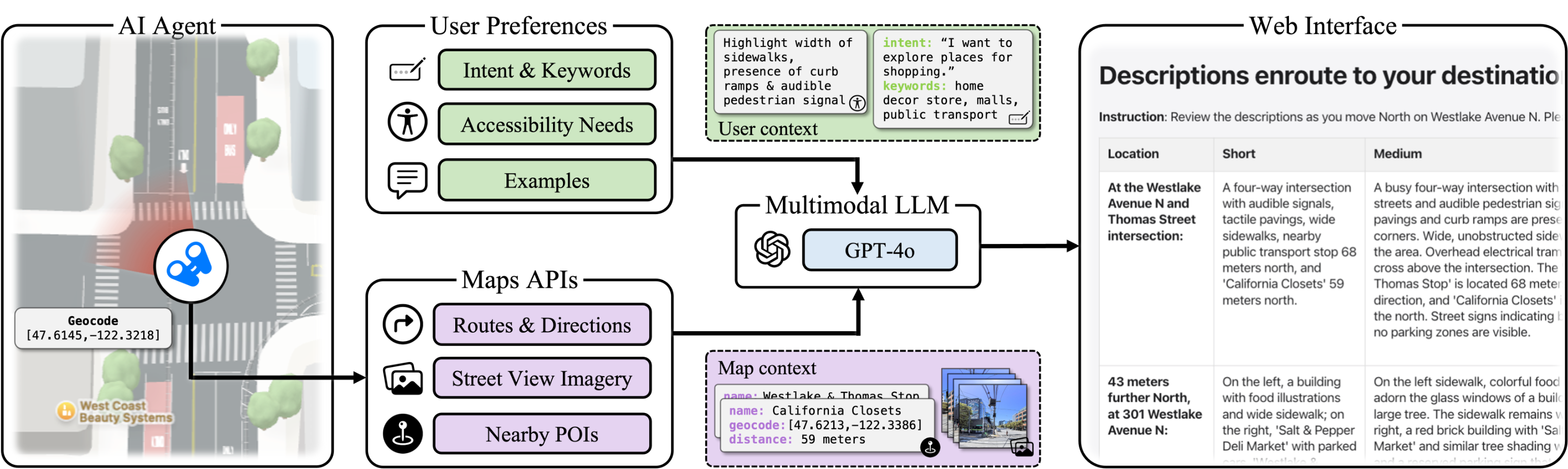}
    \caption{System architecture of \system, an MLLM-driven AI agent for accessible street view interactions. The agent grounds itself in the real world using geographic coordinates (\emph{i.e.}, geocodes) and retrieves street view imagery, routes, and POI data via \maps. BLV users’ preferences—such as intent and accessibility needs—are processed alongside map data using GPT-4o, enabling \system\ to generate textual descriptions. The web interface presents this information to BLV users.}
    \Description{SceneScout’s system architecture with five core components interacting to generate accessible street view descriptions. On the left, the AI agent is shown navigating a street intersection using geographic coordinates. Below it, the Maps APIs provide routes, street view imagery, and nearby points of interest. At the top right, User Preferences--including intent, accessibility needs, and example keywords--are defined by the BLV user. These preferences and map data are both input into a central Multimodal Large Language Model (GPT-4o). The model synthesizes this information to produce grounded textual descriptions of the environment. On the far right, the Web Interface displays the output, presenting location-based descriptions in short and medium formats. Arrows illustrate the flow of data from the AI agent and user preferences through the model to the final output, enabling a personalized, accessible navigation experience.}
    \Description{Bar chart reporting average participant ratings for perceived relevance and usefulness of two interaction modes: Route Preview (green) and Virtual Exploration (purple). Each category has two horizontal bars, with Virtual Exploration rated higher for relevance, and both modes rated similarly for usefulness. Exact mean values are included in the main text.}
    \label{fig:system_architecture}
\end{figure*}

\subsection{Implementation Details}
\label{sec:implementation_details}
\system\ is powered by a MLLM-driven AI agent that perceives the environment through street view imagery, and reasons about BLV users' preferences and context to facilitate accessible interactions with that imagery.  

Figure~\ref{fig:system_architecture} illustrates \system's system architecture. The agent positions itself by using geographic coordinates (i.e., longitude, latitude). With these coordinates, it leverages \maps\ to retrieve street view imagery, determine valid movements within the street view, gather information about nearby POIs, and plan routes. These data, along with user preferences expressed in natural language, are processed by OpenAI’s GPT-4o multimodal LLM~\cite{openai_gpt4o}, which services prompt requests. To enhance the AI agent's reasoning capabilities, 
user preferences and contextual information are encoded using few-shot prompting~\cite{brown_language_2020}, chain-of-thought (CoT) reasoning~\cite{wei_cot_2024}, and prompt-chaining techniques~\cite{wu_promptchainer_2022, yang_v-irl_2024}. Appendix~\ref{sec:prompts} includes full prompt details.

\system's web interface is implemented using HTML, CSS, and JavaScript, with a Flask server managing back-end communication with the agent. We adhered to the World Wide Web Consortium (W3C) accessibility guidelines~\cite{w3c} and tested compatibility with screen readers like VoiceOver~\cite{voiceover}. Additionally, the interface incorporates W3C-recommended table structures, including two headers, to improve usability for screen reader users~\cite{w3c_tables}.




\section{User Study Method}

\label{sec:user_eval}
Our user study is aimed at understanding BLV users' experiences interacting with \system\ to access street view imagery. To this end, we conducted a mixed-methods study with 10 BLV participants, answering the following research questions:
\begin{itemize}
    \item[\textbf{RQ1.}] How do BLV users perceive the relevance and utility of information surfaced from street view imagery by \system?
    \item[\textbf{RQ2.}] What are BLV users' attitudes toward MLLM-generated descriptions of street view imagery, particularly regarding their trust and confidence in the descriptions?
    \item[\textbf{RQ3.}] What are BLV users' usage patterns with \system\ and plans for incorporating it into their navigation workflows?
    \item[\textbf{RQ4.}] What future interactions and use case scenarios could emerge for navigation systems that leverage street view imagery beyond the scope of \system?
\end{itemize}

\subsection{Participants}
We recruited 10 screen-reader users (six males, three females, and one non-binary; aged 31--60) for the study. Participants worked at a large US technology company and were recruited by posting to internal communication channels and via snowball sampling~\cite{goodman_snowball_1961}. 

Table 1 (Appendix~\ref{sec:ptcpt_demographics}) summarizes our participants' demographics, onset of vision impairment, use of mobility aids, and proficiency in using assistive technology (AT) for navigation. All participants, except P4, were totally blind and used mobility aids such as white canes (P1, P2, P8), guide dogs (P5, P6, P7, P9), or both (P3, P10). P4 reported having low vision due to severe central-vision distortion and did not use any mobility aids. All but one participant (P6) reported themselves as being moderately--extremely proficient in using AT for navigation (3+ scores on a 5-point rating scale). Half of the participants were blind since birth, while the other half developed vision impairment later in life. Participants held a range of roles within the company, reflecting varying levels of experience with AI and MLLMs, although overall familiarity was likely higher than the general BLV population. Roles included software engineer (P1–P5, P8), business process analyst (P6), product manager (P7), data annotator (P9), and retail advisor (P10).

Study sessions were conducted remotely over video conferencing software and lasted for 90-120 minutes. Participants received \$12 meal vouchers as incentives for participation. We obtained informed consent from all participants.

\subsection{Experimental Design}
Our mixed-methods study followed a scenario-based design~\cite{rosson2002scenario}, where participants interacted with \system's two interaction modes.
Each participant experienced two scenarios per mode based on use cases for street view imagery that our blind colleagues mentioned during design discussions (Section~\ref{sec:design_rationale}):
\begin{itemize}
    \item \textbf{Route Preview}: Each scenario previewed a route, which began either at a POI or the corner of a street intersection and ended at another POI (\emph{e.g.} transit stops, restaurants, stores). Start and end points were randomly selected to lie within a 5-10 minute walking distance.
    \item \textbf{Virtual Exploration}: Each scenario involved traversing through street view imagery with a specific randomized intent such as scoping out a new neighborhood for potential relocation, discovering new restaurants or activities, and identifying lively streets to explore on foot. The scenario began at a randomly selected intersection, and lasted for 10 minutes.
\end{itemize}

For both interaction modes, participants experienced two scenarios at different locations, one familiar and one unfamiliar. We intentionally started with a familiar location specified by the participant,  typically a neighborhood where they had lived or worked. This lets us naturally onboard participants while assessing \system.  Unfamiliar locations were randomly assigned from downtown areas in major US cities. We deliberately varied participants' familiarity with the environment to gain insights into the perceived accuracy (RQ1), trust (RQ2), and utility of MLLM-generated street view descriptions (RQ3).

\subsection{Procedure}
We began each session by administering a \textit{pre-study questionnaire} (Appendix~\ref{sec:pre-study-ques}) to collect demographic information and details about participants' level of vision, use of assistive technology for navigation, and familiarity with street view imagery. Participants then interacted with one of the two interaction modes in counterbalanced order to reduce potential ordering biases.

For each mode, we first gave participants a quick tutorial on how to navigate the web interface and interpret descriptions. Participants engaged with the first interaction mode in two scenarios, during which we asked them to think aloud
~\cite{ericsson1980verbal}. After completing both scenarios, we administered a \textit{post-prototype questionnaire} (Appendix~\ref{sec:post-prototype-ques}) designed to answer RQ1--3. For RQ1 (relevance and utility of descriptions), participants rated perceived relevance and utility of descriptions on a scale of 1--5. For RQ2 (attitudes toward MLLM-generated descriptions), they rated perceived trust and confidence in the descriptions (scale 1--5), and commented on the accuracy of descriptions for familiar locations. For RQ3 (usage patterns and integration into navigation workflows), participants ranked preferred verbosity levels, rated how location familiarity influenced their use (scale 1--5), and reflected on how \system\ would fit into their current workflows. We followed up on participants' ratings with additional questions to better understand their perspectives. The same questionnaire was administered after experiencing \system's second interaction mode.

Following their experience with \system's two interaction modes, we conducted a \textit{semi-structured interview} (Appendix~\ref{sec:post-study-ques}) designed to answer RQ4 (future use cases for street view imagery-based navigation systems). The interview included open-ended questions that solicit suggestions for improving \system's two interaction modes and potential use cases for street view imagery beyond these modes. All questionnaires and interview questions are included in Appendix~\ref{sec:study-ques}.

\subsection{Analysis}
To analyze the interviews, we transcribed the study sessions in full and performed thematic analysis~\cite{braun_using_2006}. The first author led the analysis by reading the interview transcripts, taking notes, and synthesizing an initial set of codes which was shared with the other authors, who provided feedback on the initial set. From this feedback and continued iteration, the author refined codes to identify emerging themes grouped according to the research questions (RQ1--4).

\section{User Study Results}
\label{sec: user_eval_results}
In the following sections, we report themes that emerged for each of the research questions RQ1--4.


\subsection{Relevance and Utility of Descriptions (RQ1)}
\label{sec:RQ1-utility_of_descriptions}
\begin{figure} 
    \centering
    \includegraphics[width=0.85\linewidth]{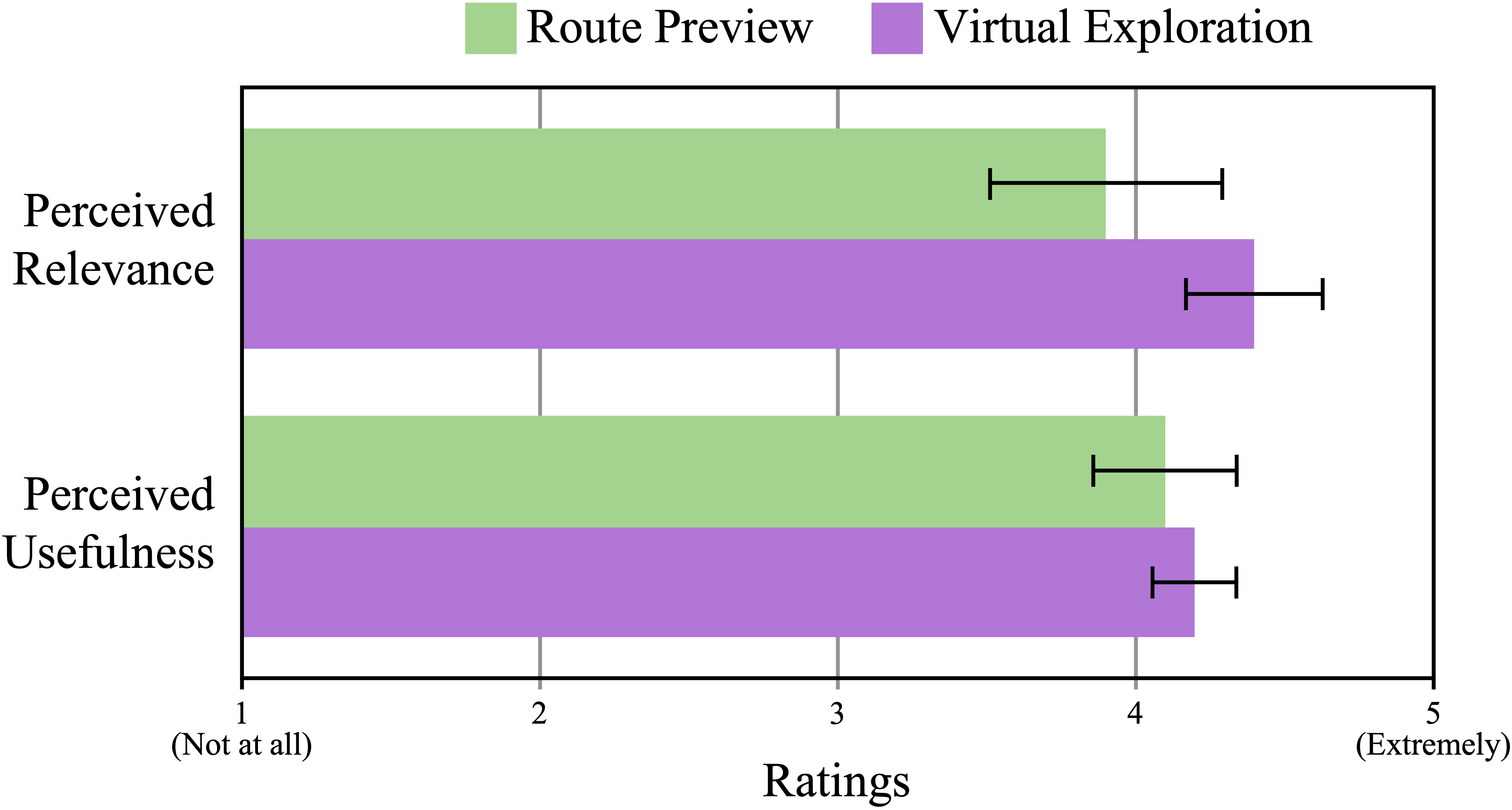}
    \caption{Participants' average ratings ($N=10$) for perceived relevance and usefulness of descriptions from \system's two interaction modes. While both modes received positive ratings, descriptions from Virtual Exploration were found to be slightly more relevant and useful compared to those from Route Previews. Error bars indicate standard error.}
    \label{fig:ratings-relevance_usefulness}
\end{figure}

Figure~\ref{fig:ratings-relevance_usefulness} shows participants' average ratings for relevance and utility of descriptions across both interaction modes. The mean ($\pm$ std. dev.) rating for participants' perceived relevance of descriptions was $3.9$ ($\pm 1.2$) for Route Preview and $4.4$ ($\pm 0.7$) for Virtual Exploration. The mean ($\pm$ std. dev.) rating for participants' perceived utility of descriptions was $4.1$ ($\pm 0.7$) for Route Preview and $4.2$ ($\pm 0.4$) for Virtual Exploration. These positive ratings suggest that participants valued the information surfaced from street view imagery, instilling a sense of independence. P1 echoed this sentiment: \textit{``these are details that a sighted person may not find useful to describe, but a blind person might not know how to ask.''} Follow-up discussions revealed the following key themes on relevance and utility of descriptions.


\subsubsection{\textbf{Useful categories of visual information.}}
\label{sec:results-category_info}
Participants found a variety of visual details useful for route previews, including landmarks like transit stops and businesses, accessibility-related features (\emph{e.g.}, curb cuts, tactile pavings, APSs), and traffic control elements (\emph{e.g.}, stop signs, lights, one-way vs. two-way streets). They also valued information about intersection types, infrastructure features (\emph{e.g.}, sidewalks, surface textures, pedestrian crossings), signboard text, mobility cues or obstacles (\emph{e.g.}, trash cans, benches, fire hydrants, bike racks, planters), and potential auditory cues like fountains. Some appreciated mentions of minute details, such as tripping hazards like tree roots or sidewalk cracks, which are difficult to detect otherwise. City-specific structures like Minnesota’s skywalks were especially useful, as shown in Figure~\ref{fig:table_with_reactions}. P7 noted, knowing about such features ahead of time  \textit{“shortens the learning curve significantly,”} reducing the need to rely on others when navigating new areas.

In virtual exploration, participants valued the richer descriptions, which helped them \textit{``paint a fairly decent picture of what’s going on in the neighborhood''} (P3). As P4 described, \textit{``glass facades sets all the buildings up around you.''} However, the subjective and sometimes vague language in these descriptions caused confusion. P1, for example, found non-specific terms frustrating, stating, \textit{“Avoid terms like `landscaped area.' It’s either grass, dirt, or a park... we should get it to avoid extremely vague things.”}


\subsubsection{\textbf{Inferring environment's affordances.}} 
\label{sec:results-env_affordances}
In addition to aiding navigation, participants stated the descriptions influenced their real-world decision-making by revealing aspects of the environment not apparent from navigational cues. For example, P7 appreciated the mention of an \textit{`Only Authorized Vehicles'} sign:
\begin{quote}
\textit{``If I didn't know this area, I wouldn't know that only buses and emergency vehicles are allowed on Nicollet Avenue. It's interesting because if you were trying to get an Uber, you wouldn't catch it there.''} --\textbf{P7}
\end{quote}
Participants also appreciated descriptions that conveyed contextual details about public infrastructure. For instance, learning that a bus shelter is transparent can be especially useful in bad weather, as it allows riders to remain sheltered while still being visible to the driver---unlike opaque shelters that may require stepping out to signal the bus. These kinds of details helped participants adapt their behavior and make more informed decisions, even if the information was not strictly related to navigation.


\begin{figure}
    \centering
    \includegraphics[width=0.85\linewidth]{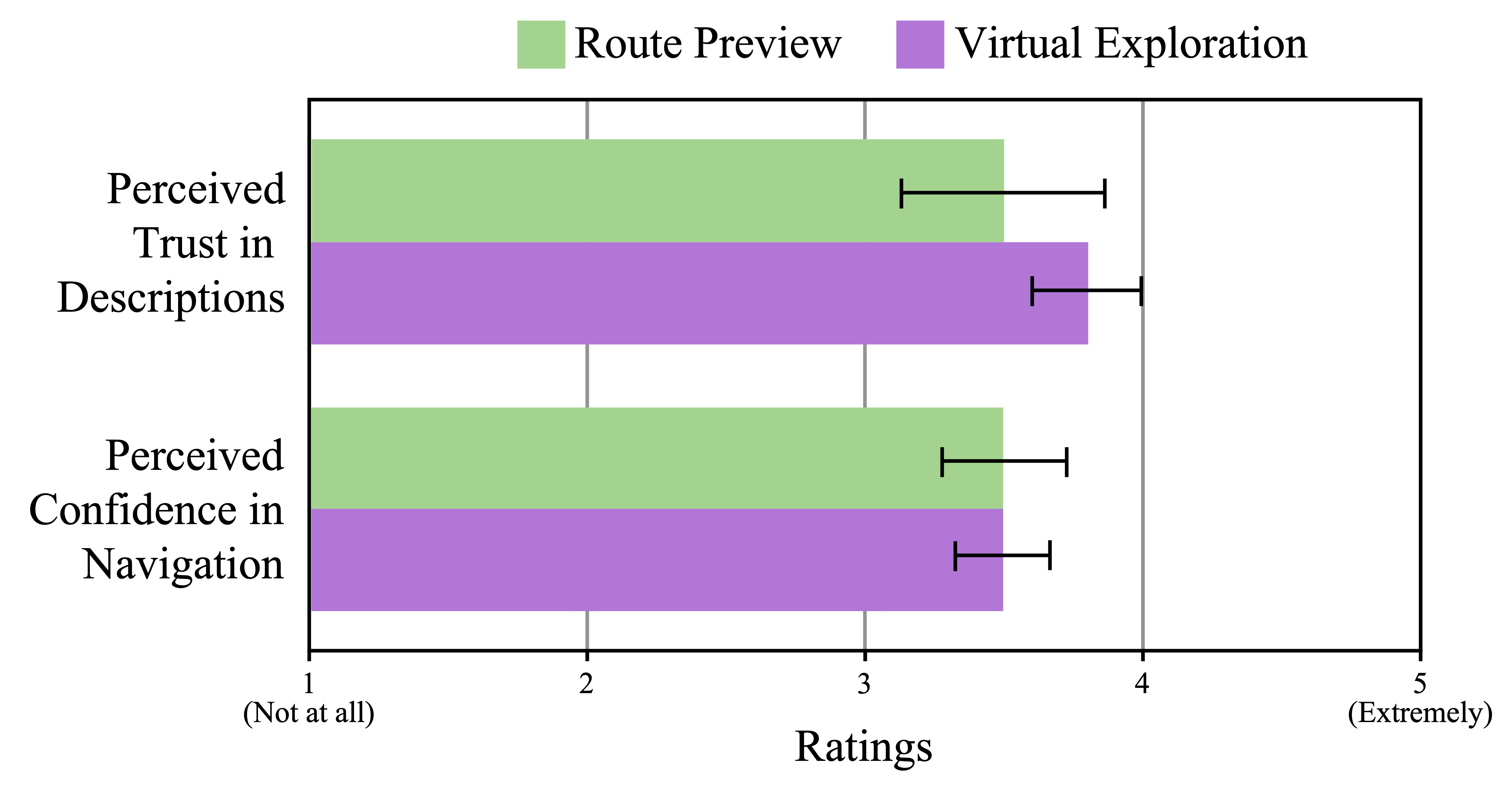}
    \caption{Participants’ average ratings ($N=10$) for perceived trust in descriptions and confidence in navigation across \system's two interaction modes. Descriptions from Virtual Exploration were trusted slightly more than those from Route Preview. Both modes instilled a similar level of confidence in participants to navigate based on the information provided. Error bars indicate standard error.}
    \Description{Bar chart reporting average participant ratings for perceived trust in descriptions and confidence in navigation for Route Preview (green) and Virtual Exploration (purple). Each category has two horizontal bars, with Route Preview slightly lower on trust but exactly the same for confidence. Exact mean values are included in the main text.}
    \label{fig:ratings-trust_confidence}
\end{figure}

\subsubsection{\textbf{Specifying keywords enhances relevance.}}
\label{sec:results-keyword_control}
Participants appreciated the ability to specify keywords in virtual exploration, giving them control to steer the system to focus on details that mattered the most to them. P6 explained how this customization could align the descriptions better with her routine, shifting based on whether she was alone or with her partner:
\textit{``I know it said public transport, [but] I don’t love the subway system in New York. I'd rather focus on using buses... and I wanted to learn about the parking spaces there because my partner drives''} (P6).  However, some participants desired to further refine the descriptions by not only adding but also modifying or removing suggested keywords, .

\subsubsection{\textbf{Route preview descriptions lacked spatial precision.}}
\label{sec:results-precise_spatial_info}
Participants found descriptions slightly less relevant and useful for route previews than for virtual exploration (Figure~\ref{fig:ratings-relevance_usefulness}) because route information required greater precision to support effective navigation. 
Participants emphasized the importance of describing object locations relative to their path. P1 noted that instead of just stating the presence of curb cuts, descriptions should clarify their layout: \textit{``In [my city], some intersections have curb cuts that are at $90^\circ$ angles, while the other side have them at $45^\circ$. A helpful description might be: You're at a four-way intersection, and at the southeast corner are two curb cuts at $90^\circ$; the northwest corner is a single curb cut at $45^\circ$.''} Exploration-focused descriptions could afford to be broader, providing a general sense of the environment rather than precise navigational details. As P3 noted: \textit{``For exploration, the more verbose descriptions are useful since I don't have a specific goal in mind. I'm trying to understand what I would experience there. For exploring, the descriptions are pretty darn good.''}

\subsection{Attitudes Toward MLLM-generated Street View Descriptions (RQ2)}
\label{sec:RQ2-attitudes}
Figure~\ref{fig:ratings-trust_confidence} shows participants' average ratings for their attitudes toward MLLM-generated descriptions for both interaction modes. The mean ($\pm$ std. dev.) rating for perceived trust in the descriptions was $3.5$ ($\pm 1.2$) for route previews and $3.5$ ($\pm 0.7$) for virtual exploration. The mean ($\pm$ std. dev.) rating for perceived confidence in knowing what to expect from the environment during navigation was $3.8$ ($\pm 0.6$) for route previews and $3.5$ ($\pm 0.5$) for virtual exploration. Although participants' ratings leaned positive, they expressed hesitation in fully trusting the descriptions, acknowledging that \textit{``AI can hallucinate, let's be honest''} (P9). Figure~\ref{fig:table_with_reactions} provides examples of both errors and useful information encountered by the participants during the study.

\subsubsection{\textbf{Trust through familiarity and physical verification.}}
Participants gauged the trustworthiness of MLLM-generated street view descriptions by evaluating their alignment with prior knowledge, real-world verification, and consistency over time. Descriptions that matched familiar environments were perceived as more reliable. For example, P5 felt reassured when a description of Central Avenue reflected their expectations: \textit{``It mentions that going north, there's gonna be some apartment buildings with ground-level shops, and that is exactly correct.''} However, participants emphasized that building trust required time and repeated interactions with the system. As P5 further explained: \textit{``I would say a four [out of five], because they seem trustworthy for the areas I was familiar with, but my sample size isn’t big enough.''}

In addition to familiarity, participants stressed the importance of physical verification. P7 described \system\ as \textit{``a really useful tool in my toolbox, but it doesn’t feel like I can say exactly how the environment is going to be.''} P4 echoed this skepticism, stating: \textit{``I don't trust nothing until I touch that pedestrian signal. It could not be there. It could be there. It could be broken... but it gives you an expectation of what should be there.''}

\begin{figure}[t]
    \centering
    \includegraphics[width=\linewidth]{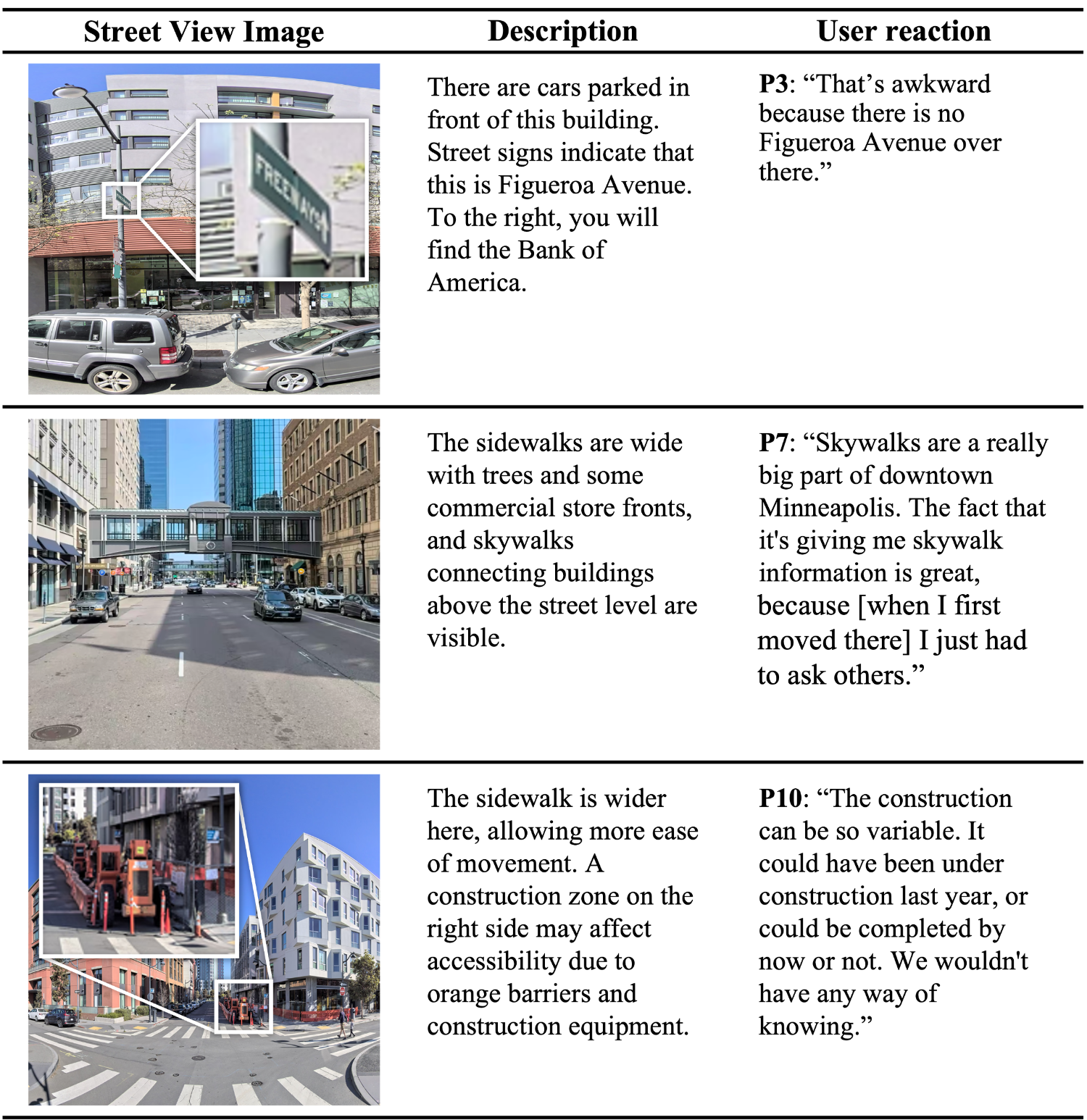}
    \caption{
    Examples of errors and useful information encountered by the participants during the study. The first highlights a misreading of a signboard, leading to incorrect location information. The second demonstrates the usefulness of describing skywalks, which are otherwise inaccessible details for BLV users. Third shows a temporally inconsistent description of a construction zone, illustrating challenges in conveying dynamic environmental changes.}
    \Description{Three-row table showing street view images, AI-generated descriptions, and participant reactions. Each row includes a photo on the left, a corresponding textual description in the center, and a user quote on the right. The first row shows a street scene with a zoomed-in sign labeled “Figueroa Ave.” The description references the street name and a nearby Bank of America, but the participant reacts with confusion, noting the location is incorrect. The second row displays a downtown street with elevated skywalks between buildings. The description highlights wide sidewalks and skywalks, and the participant appreciates the inclusion of this information, which had previously been hard to access. The third row shows a sidewalk with a construction zone marked by orange barriers. The description notes improved sidewalk width and reduced accessibility due to the construction. The participant comments on the temporal uncertainty of such scenes, pointing out that the construction status could have changed since the image was taken.}
    \label{fig:table_with_reactions}
\end{figure}

As shown in Figure~\ref{fig:table_with_reactions}, inaccurate descriptions undermined participants’ confidence. P4, for instance, encountered a hallucinated street name and immediately questioned its validity:\textit{``I don't know if that's actually true. [The] thing that is further north on 4th Street from Mission Bay Blvd., the next possible street is, oh wait a minute, that's awkward because there is no Figueroa Ave. over there.''} Likewise, P3 expressed frustration when a description incorrectly stated that audio pedestrian signals were present: \textit{``If the data is incorrect, I'm not going to trust it at all.''} Participants also highlighted the challenge of temporal inconsistencies, such as transient objects like vehicles and construction. More details   are in Appendix~\ref{sec:results-temporal_consistency}. 


\subsubsection{\textbf{Assumptions about people's abilities and actions can be offensive.}}
Participants expressed dissatisfaction with the prescriptive nature of the descriptions, emphasizing the need for greater objectivity and avoidance of assumptions about users' abilities. Upon encountering the description: \textit{`...the sidewalks are wide, but caution is advised due to frequent obstructions, including parked vehicles,'} P8 exclaimed, \textit{``That sentence doesn't need to be included. Don't tell me what I should do. Tell me the information.''}
Similarly, participants highlighted the importance of refraining from making assumptions about users' actions:
\begin{quote}
    \textit{``The fact is, we don't know how a user is going to cross the street. Are they gonna use another app to tell them the walk sign is on? Are they going to facetime a friend? We shouldn't make assumptions. I would be cautious about editorializing.''} --\textbf{P1}
\end{quote} 
Additionally, participants advised against explicitly naming disabilities in descriptions; for instance, referring to tactile pavings as meant for `blind people' should be avoided.

\subsubsection{\textbf{Assumptions about places can be misleading.}}
Participants pointed out that some descriptions made unwarranted assumptions about places without sufficient evidence. For instance, when exploring new neighborhoods for relocation, descriptions sometimes inferred that an area was quiet without providing concrete facts. P4 felt that such assumptions should be avoided: 
\begin{quote} \textit{``How does it know that it's a quiet residential area? I almost think it's making that up. It can set somebody up for false expectations. Being objective is the best thing, and then allow the user to determine the implications.''} --\textbf{P4}
\end{quote}

\subsection{Usage Patterns and Integration into Navigation Workflows (RQ3)}
\label{sec:RQ3-nav_workflow_integration}

\subsubsection{\textbf{Usage of different verbosity levels.}}
\label{sec:results-verbosity_levels}
Participants preferred short and medium descriptions for Route Previews and medium and long descriptions for Virtual Exploration. Figure~\ref{fig:rankings_descriptions} (Appendix~\ref{sec:ranking_pref_appendix}) presents their ranking preferences for different verbosity levels, although all participants valued the ability to switch between verbosity levels as needed. For example, P8 said \textit{``all three [levels] have merits''} (P8), while P6 explained, \textit{``I like [...] the medium, and then if I want more details, I'm going to the long.''} 
When in a hurry, short descriptions were preferred because \textit{``it gets straight to the point''} (P9). Familiarity also played a role, as P4 noted, \textit{``If I'm in familiar [location], I would probably use the short. If I'm in unfamiliar, I would want the medium or the long to learn and understand the area.''} Visit purpose further shaped preferences, as P6 shared, \textit{``If I'm visiting for a week and I just need to get a smoothie, then I would use medium or short. I just want to get in, get my thing, and get going. I won't come back and don't need descriptions of anything around the area.''}

\subsubsection{\textbf{Effect of location familiarity on system usage.}}
\label{sec:results-familiarity_effect}
Prior research suggests that pre-travel navigation assistance is primarily useful in unfamiliar locations~\cite{kameswaran_understanding_2020, guerreiro_virtual_2017, guerreiro_virtual_2020}. However, we found that the level of detail available in street view imagery makes it valuable even in familiar areas.

\begin{figure}[t]
    \centering
    \includegraphics[width=0.85\linewidth]{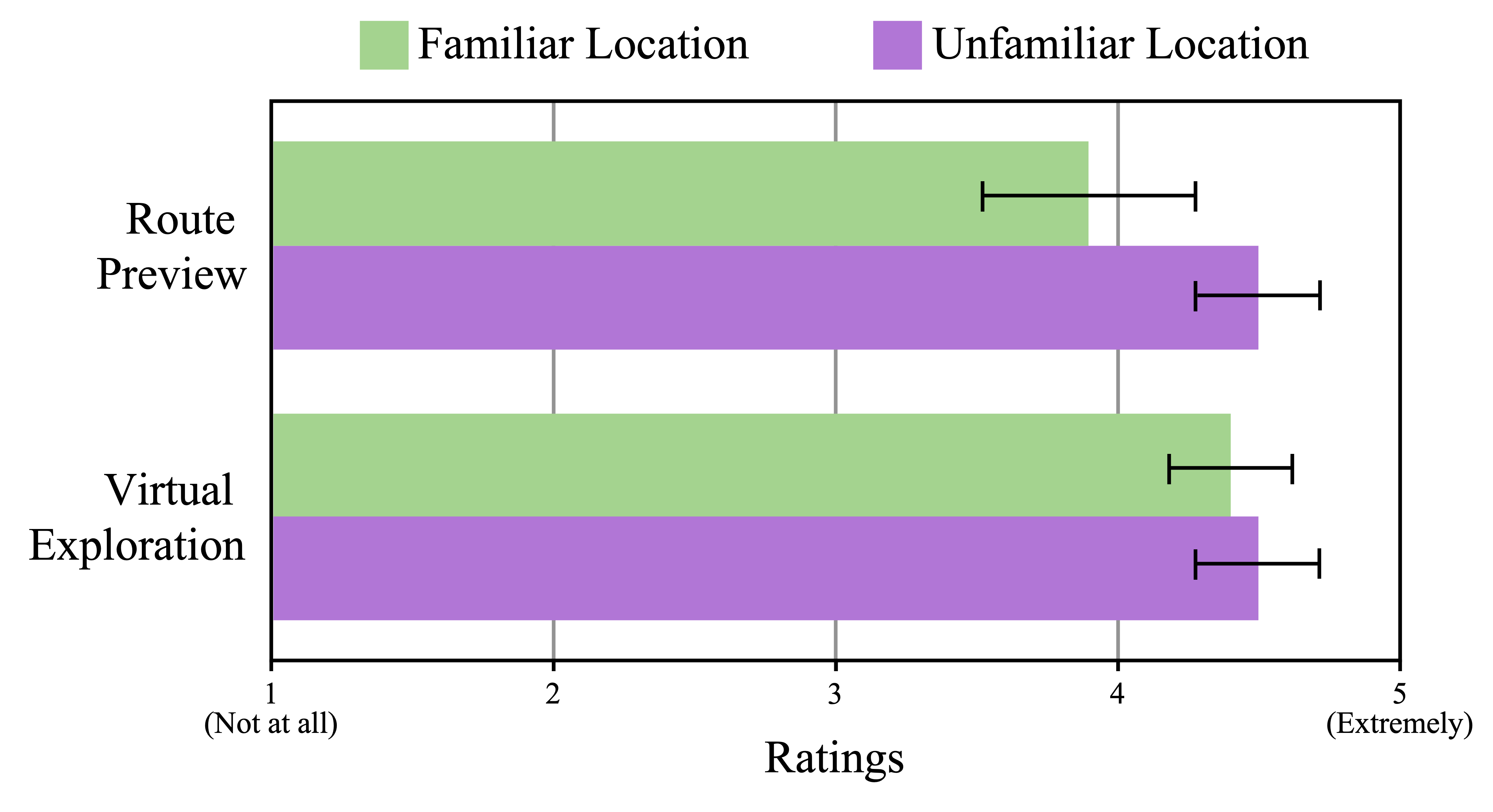}
    \caption{Participants’ average ratings ($N = 10$) for their likelihood of using \system’s two modes in familiar and unfamiliar locations. While participants were equally likely to use Virtual Exploration regardless of location familiarity, they were slightly more likely to use Route Preview in unfamiliar locations. Error bars indicate standard error.}
    \label{fig:ratings-familiarity_effect}
    \Description{Bar chart reporting average participant ratings for likelihood of using SceneScout’s interaction modes in familiar (green) and unfamiliar (purple) locations. Each mode—Route Preview and Virtual Exploration—has two horizontal bars. Ratings for unfamiliar locations are higher across both modes. Exact mean values are included in the main text.}
\end{figure}

Figure~\ref{fig:ratings-familiarity_effect} presents participants’ average ratings for their likelihood of using the two interaction modes in both familiar and unfamiliar locations. The mean likelihood ($\pm$ std. dev.) for Route Preview was $3.9$ ($\pm 1.2$) in familiar locations and 4.5 ($\pm 0.7$) in unfamiliar locations. For Virtual Exploration, the ratings were $4.4$ ($\pm 0.7$) and $4.5$ ($\pm 0.7$), respectively. Participants expressed a strong likelihood of using both modes regardless of location familiarity, particularly for Virtual Exploration.

In unfamiliar locations, participants valued both modes for building confidence and setting expectations. As P5 noted: \textit{``This would be really useful for people who want more information about what to expect because blind people have varying degrees of travel skills and varying degrees of comfort with unfamiliar environments. Having a better idea of what's coming could be a really big help.''}

Contrary to expectations, participants also found Route Preview valuable in familiar locations. As P10 explained, they used it \textit{``to re-familiarize myself with [the area] to make sure that I'm not forgetting anything.''} P8 echoed this, acknowledging that even in well-known areas, one can still miss things or not know that something exists. P6 illustrated this with an example: \textit{``I'm not gonna lie, I know that sidewalk, but I didn't know the fence [behind] was wrought iron.''}

Similarly, Virtual Exploration in familiar locations offered context that was otherwise inaccessible. P1 highlighted this gap:  \begin{quote}
\textit{``Unless you have a very specific friend, there are pieces of your world that no one might tell you about. They may not tell you there's a mural there. Maybe I've lived here for ten years and I happen to be with a friend from out of town, and they're like, `oh, that's a pretty mural.' I'm like, what mural?''} --\textbf{P1}
\end{quote}

\subsubsection{\textbf{Incorporating \system\ into navigation workflows.}}
\label{sec:results-nav_workflows}
Participants identified several use cases for incorporating \system\ into their navigation workflows. First, they wanted to establish a common visual ground with sighted people. Using visual references, for example, could enhance communication clarity:
\begin{quote}
\textit{``I love the description of color! I mean, you would never know that there's a purple umbrella, but then if someone was looking for you, you could say: `Hey, I'm by the purple umbrella on 17th and Pearl Street,' you know? These are really powerful descriptions for different use cases.''} --\textbf{P8}
\end{quote}
Additionally, participants emphasized that visual information enhances their ability to \textit{``ask a better question of somebody who's around''} (P7).


Second, participants highlighted how \system\ could help to build mental maps, fostering confidence in navigating new areas:
\begin{quote}
\textit{``Some blind people find it a little difficult to explore and are focused on getting from point A to point B. They don't necessarily know a lot about what is along the way. So, this [system] really helps with getting a more full picture of what's there and engage more with your environment.''} --\textbf{P5}
\end{quote}
Another participant (P9) noted newfound confidence in exploring an area they had previously only traversed by bus: \textit{``I haven't actually been here on foot because I didn't know it has sidewalks that aren't that bad. I think I could walk in.''}

Third, participants expressed excitement about the Virtual Exploration mode. P8 said: \textit{``I would sit there and play with the [system] along Pearl Street as long as you would let me.''} They identified various potential uses, including visiting their hometowns (P1, P6, P7, P8, P9, P10), planning vacations (P1, P6, P10), following up on newsworthy locations or friends' new homes (P1, P6, P10), exploring new neighborhoods or workplaces (P1, P6, P7, P8), and assessing locations for business meetings or outings with friends (P2, P4, P6).

\subsection{Ideas for Improvement and New Interactions (RQ4)}
\label{sec:RQ4-future_interactions}

\subsubsection{\textbf{Suggestions for improvement.}}
Participants suggested several improvements to \system. First, they wanted personalized descriptions that adapt over multiple sessions rather than relying solely on single-session keywords. P8 proposed \textit{``being able to mark certain elements of descriptions as favorites, so the model knows that [a user] likes to hear about these things.''} 
Second, participants recommended shifting the point of view from a vehicle’s perspective to a pedestrian’s. P3 explained that interpreting car-centered imagery is challenging: \textit{``It has to be interpolated as if you were walking up the sidewalk rather than being in the middle of the street. People who are blind do not think about the street in a fully spatial way, so you can't just say this is from the point of view of the center of the street.''} 
Finally, they emphasized the need for better exploration capabilities so that users can move beyond the current window and retrace their steps. P8 pointed out that blind users often lack the same freedom to explore and may need to backtrack to find alternatives.

\subsubsection{\textbf{Integration into walking directions for in-situ assistance.}}
Participants expressed a strong desire for real-time access to street view descriptions while walking. They envisioned applications that surface visual information through bone conduction headphones or transparency mode to provide relevant details as they move. As P9 put it, \textit{``Why can't [maps] have a built-in ability to help [provide] detailed information about what you're walking by.''}

Participants suggested using even shorter, \textit{`mini'} (P1), descriptions while walking, highlighting only critical details such as landmarks or sidewalk conditions. More comprehensive descriptions, \emph{i.e.} long descriptions, could be triggered on demand when users pause walking or reach intersections. 

Another participant (P4) suggested a new form of interaction, in which users \textit{``could point the device in a certain direction''} to receive on-demand descriptions, rather than having to physically align their phone camera to capture the surroundings. This would enable users to actively survey their environment in real time, making navigation more dynamic and responsive.

\section{Technical Evaluation}
\label{sec:tech_eval}

\begin{figure*}[t]
    \centering
    \includegraphics[width=0.99\linewidth]{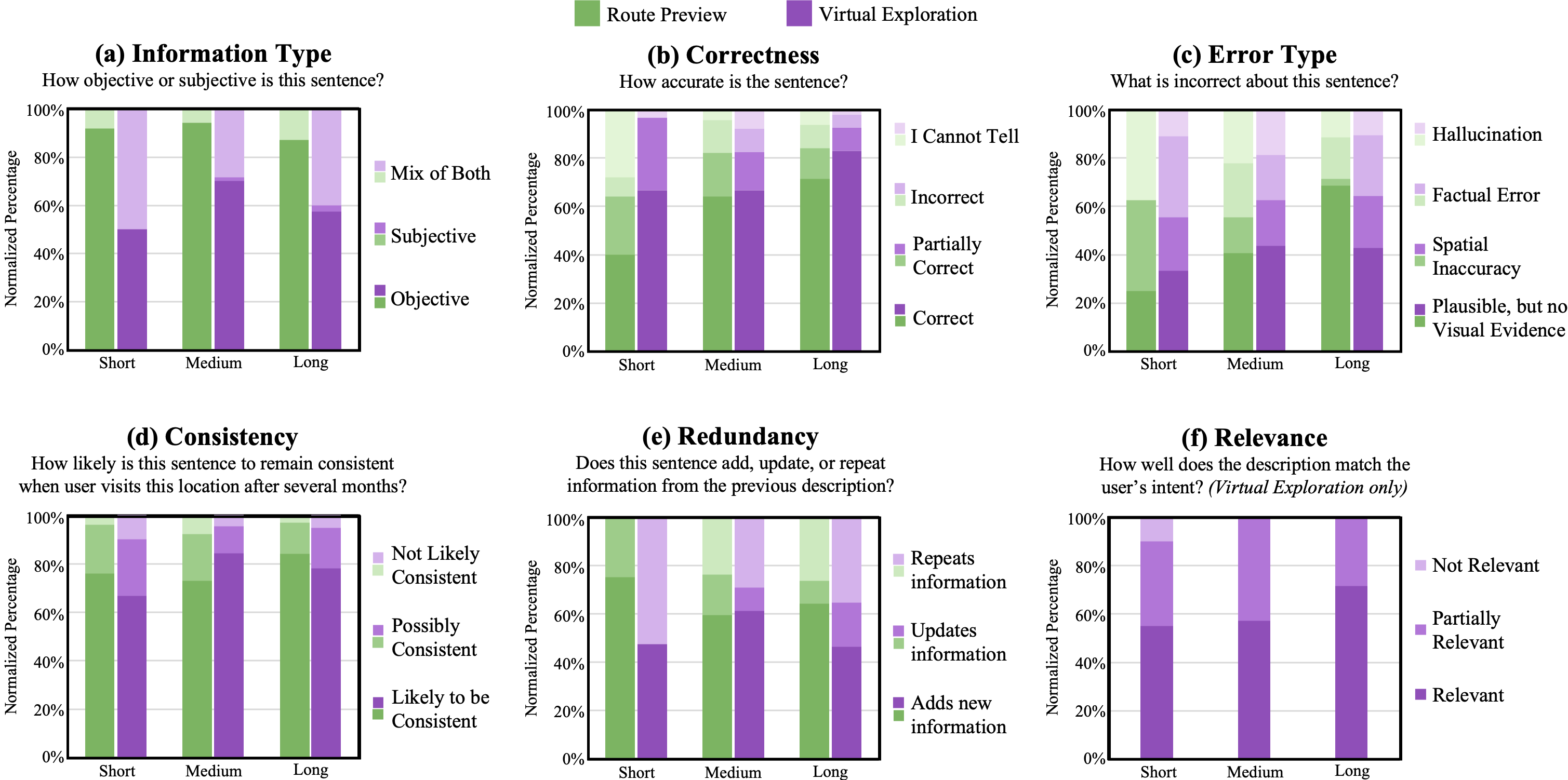}
    \caption{Technical evaluation results for (a)~information type, (b)~correctness, (c)~error type, (d)~temporal consistency, (e)~redundancy, and (f)~relevance of descriptions. We report percentages comparing both of the interaction modes as well as description length. All of the metrics except relevance are analyzed on a per-sentence basis (\emph{i.e.}, results are normalized for description length). We find that while most descriptions are accurate and reliable, the errors are often subtle and plausible, making them difficult to detect without seeing the images.} 
    \label{fig:tech_eval_results}
    \Description{Six grouped bar charts reporting technical evaluation results of SceneScout descriptions across Route Preview (green) and Virtual Exploration (purple), analyzed by description length (short, medium, long). (a) Information Type shows the mix of objective, subjective, and mixed statements. (b) Correctness rates descriptions as correct, partially correct, incorrect, or uncertain. (c) Error Type categorizes errors into hallucination, factual error, spatial inaccuracy, or plausible but unverifiable content. (d) Consistency evaluates how likely descriptions remain valid over time. (e) Redundancy measures whether sentences add, repeat, or update information. (f) Relevance (Virtual Exploration only) shows how well descriptions align with the user’s intent. Each bar is normalized to 100\%, and percentages are reported in the main text.}
\end{figure*}

We evaluate \system's technical performance to determine how accurately and reliably the MLLM extracts information from street view imagery. Our user evaluations (Section~\ref{sec:user_eval}–\ref{sec: user_eval_results}), which captured BLV users’ \textit{subjective} experiences with \system, revealed instances of hallucinations, incorrect assumptions, and irrelevant details in street view descriptions. To further examine these issues, our technical evaluation provides an \textit{objective} analysis of \system’s performance. Specifically, we assess the generated descriptions for information type, correctness, error type, consistency, redundancy, and relevance.

\subsection{Procedure}
We selected a set of $40$ usage logs from \system\ for evaluation, evenly split between the two interaction modes ($10$ participants $\times$ $2$ scenarios $\times$ $2$ interaction modes). Overall, the dataset covered downtown areas in Austin, Chicago, Los Angeles, Minneapolis, New York City, Pittsburgh, San Francisco, Seattle, and Sunnyvale. We randomly sampled $20\%$ of the descriptions from these logs for the technical evaluation, resulting in $126$ descriptions. The analyzed descriptions had average lengths ($\pm$ std. dev.) of $1.46$ ($\pm0.7$) sentences for short descriptions, $3.57$ ($\pm1.3$) sentences for medium descriptions, and $7.95$ ($\pm1.8$) sentences for long descriptions. In total, across the two interaction modes and three verbosity levels, we evaluated over $550$ sentences.

Each sentence was evaluated across five metrics: \textit{information type, correctness, consistency, redundancy}, and \textit{relevance}. Note that we only rated \textit{relevance} for Virtual Exploration descriptions, assessing how well they matched the user's intent. The definitions for these metrics follow prior work~\cite{mohanbabu_context-aware_2024, gardner_determining_2020, huh_long-form_2024, levinboim_quality_2021}. We compared each sentence with the relevant street view imagery, contextual user data, and map information used as input to MLLM to generate descriptions. Annotation instructions for an example task are provided in Appendix~\ref{sec:annotation_instructions}. Two researchers jointly evaluated a small subset of descriptions, subsequently discussing and resolving any differences. One of the researchers then evaluated the remaining data.\looseness=-1

\subsection{Results}

Figure~\ref{fig:tech_eval_results} summarizes the technical evaluation results and compares both of the interaction modes as well as description length. All of the metrics except relevance are presented on a per-sentence basis, meaning the results are normalized for description length.

\subsubsection{\textbf{Type of Information}}
Some participants in the user study desired \system\ to refrain from making subjective statements. This evaluation showed that the majority (74\%) of information in the sentences were objective descriptions of the visuals. An additional 25\% of sentences contained both objective and subjective information, often using adjectives such as "busy", "crowded", "ample", "well-maintained", "well-kept", "quiet", or "bustling".  There were substantially more examples of this in the Virtual Exploration interaction mode, primarily when the model attempted to answer the user's intent.

\subsubsection{\textbf{Correctness and Error Types}}
Of the sentences graded, 72\% were rated correct and 14\% partially correct, with 8\% fully incorrect. Six percent were rated "I cannot tell" as the statement could not be verified  with the images. In the Route Preview mode, short descriptions have more sentences containing errors, although this is often due to these sentences being more information-dense, meaning there is more potential for a mistake in a single sentence. Across interaction modes and description lengths, the errors tended to be a fairly consistent mix of plausible but not present visual details (40\%, \emph{e.g.} stating a crossing had pedestrian signals when it did not), factual error (19\%, \emph{e.g.} incorrect text), spatial errors (16\%, \emph{e.g.} "left" instead of "right"), and outright hallucinations where the model invented whole objects (16\%, \emph{e.g.} a "Maintenance Yard" sign that was not present). Many of these errors are unlikely to be caught by BLV users unless they are familiar with the area, indicating caution must be taken when relying on MLLMs for future navigation.

\subsubsection{\textbf{Temporal Consistency}}
As noted by participants, the imagery (and therefore model responses) may be out of date either when the user views it or when the user visits the location. In the analysis, 79\% of sentences were likely to remain consistent over time as they described elements of the built environment such as buildings, streets, or foliage. For 16\% of sentences, some element made it only possible to remain consistent over time, usually because it described how busy an area was for people or vehicles, or described some seasonal element. Five percent of sentences were likely outdated because they described a temporary event (\emph{e.g.} construction) or transient object (\emph{e.g.}, a specific person or vehicle).

\subsubsection{\textbf{Redundancy}}
As users experience the descriptions while moving down a street, elements such as lane configuration, sidewalk quality, and the presence of parking are likely to be similar between descriptions. We examined the sentences in the current description with the description presented to the user immediately prior, finding that 56\% added new information not in the prior description, 31\% repeated information already told to the user, and 13\% updated information from the prior description. Redundancy was more common in Virtual Exploration as the model attempted to address the user's intention at each location, even if not much changed. Additionally, we see less redundancy in short descriptions where there is less information to repeat. 

\subsubsection{\textbf{Relevance}}
Overall, descriptions in the Virtual Exploration mode addressed the user's intention, with 61\% of descriptions being fully relevant and 35\% of descriptions being partially relevant. Partially relevant descriptions tended to also describe non-relevant aspects of the imagery, which may be desirable. Four percent of descriptions did not address the user's intention, which occurred when verbosity was limited and the model focused on what was visible in the image instead.

In summary, most descriptions were accurate (72\%), objective (74\%), likely to remain consistent over time (95\%), and provided relevant information aligned with user's intent (96\%). However, the errors that did occur were often subtle and plausible, making them difficult to detect without the ability to see the images.

\section{Discussion}
\label{sec:discussion}
We reflect on \system\ by discussing mechanisms to enhance user trust, strategies to achieve personalization at scale, the integration of map metadata with street view imagery, design considerations for street view imagery-based navigation systems, and implications for MLLMs applied to the accessibility of street view imagery.

\subsubsection*{\textbf{Designing to Enhance User's Trust.}}
Our findings revealed insights into users’ trust in AI-generated street view descriptions (Section~\ref{sec:RQ2-attitudes}). Trust significantly influenced whether BLV people would use \system-like systems, especially when they cannot physically verify those details. Participants showed cautious optimism, but remained aware of potential inaccuracies inherent in LLM-based systems~\cite{kaddour_challenges_2023, zhang_sirens_2023, huh_long-form_2024, ji_survey_2023}. Prior research explored how BLV users interpret AI-generated descriptions~\cite{adnin_i_2024, macleod_understanding_2017}, but trust becomes critical in navigation, where safety is paramount and verification is infeasible.

Two major concerns emerged: the accuracy of descriptions and the temporal consistency of imagery with real-world environments. Users need ways to verify descriptions on demand, particularly when uncertain. Existing approaches, such as touch-based interactions in ImageExplorer~\cite{lee_imageexplorer_2022} and ImageAssist~\cite{nair_imageassist_2023}, enable users to explore objects and gather objective information. Similarly, visual question answering~\cite{bigham_vizwiz_2010, gurari_vizwiz_2018, kaniwa_chitchatguide_2024} provides a way to ask context-specific questions to clarify details. Beyond user-driven verification, recent techniques that proactively detect and correct inconsistencies could further enhance reliability~\cite{chen_hallucination_2023, li_halueval_2023, gunjal_detecting_2024, manakul_selfcheckgpt_2023}.

Uncertainty cues embedded within descriptions can also help, as communicating confidence levels helps users better assess reliability. Steyvers et al.~\cite{steyvers_what_2025} found that users overestimate AI accuracy, but uncertainty phrasing aligns their confidence more closely with actual performance. 
For example, if a signboard is obscured by trees, the model could highlight this to signal potential unreliability. Additionally, providing the capture date of street view imagery could further help users assess its trustworthiness.

\subsubsection*{\textbf{Achieving Personalization at Scale.}}
Our participants prioritizing different information in street view imagery such as visual landmarks, pedestrian routes, or environmental details. They emphasized the need for personalized descriptions and greater customization, although achieving this poses challenges due to the high computational costs. Generating descriptions for each user using MLLMs is resource-intensive, making large-scale personalization with the current design of \system\ economically infeasible. As personalization is crucial for accessibility~\cite{morrison_understanding_2023, wen_find_2024, kacorri_people_2017, hong_blind_2022}, the community must balance customization with scalability~\cite{kirk_benefits_2024}.

One promising approach for achieving scale is pre-processing descriptions, enabling fast retrieval without real-time inference. However, this method offers ``one-size-fits-all'' descriptions that may not meet individuals’ specific needs. A hybrid model could refine these descriptions pre-generated by larger MLLMs using a smaller, on-device model. These lightweight models could adjust descriptions in real-time based on each user’s unique history and needs, ensuring personalization without the computational demands of large models for every query.

The key challenge is balancing scalability with the highly personalized nature of assistive technologies. Future research should optimize this trade-off to ensure that personalization at scale is effective. Emerging advancements in machine learning provide promising directions for achieving this balance~\cite{kirk_benefits_2024, tan_democratizing_2025, li_personalized_2024, salemi_lamp_2024}.

\subsubsection*{\textbf{Integrating Map Metadata with Street View Imagery.}}
\system's AI agent combines structured map metadata from \maps\ with MLLM-generated descriptions from street view imagery. Our evaluations revealed that map metadata is generally more accurate and up-to-date, but it lacks the nuanced details available in street view imagery. Metadata primarily covers POIs and driving directions, which may not always align with BLV users’ navigation needs. Additionally, walking route visuals are rarely conveyed in non-visual formats because navigation services assume users can interpret traditional map visuals, street view, or satellite images.

This gap presents an opportunity to enrich structured map metadata with the contextual information available from street view imagery~\cite{ito_zensvi_2024}. Cross-referencing both sources could provide a more comprehensive understanding of the environment, making pedestrian routes more accessible. Conversely, map metadata---given its established, reliable workflows---could help validate and refine street view descriptions, improving their accuracy and consistency. Integrating street view images into search functions within map applications could enhance search results with visual context and improve navigation for all users. Future systems should explore ways for models to distinguish between reliable map metadata and the more variable street view descriptions, ensuring that users can trust the information presented, particularly in high-stakes navigation scenarios.

\subsubsection*{\textbf{Design Considerations for Leveraging Street View Imagery.}}
Our evaluations found that street view imagery provides valuable information that BLV users would otherwise be unable to access. Currently, they often rely on friends, family, or orientation and mobility instructors for interpretation. Many participants expressed excitement about the prospect of independently accessing street view, with several echoing P7’s sentiment: \textit{“I can’t express to you how thrilled I am that I [can access] street view imagery.”} However, fully leveraging street view for accessible navigation requires addressing key design considerations.

One concern is the age of street view imagery. Prior work shows strong agreement between virtual audits of pedestrian infrastructure and traditional methods~\cite{wilson_assessing_2012, badland_can_2010, rundle_using_2011, hara_improving_2015, hara_tohme_2014}, demonstrating that even older images can be reliable. However, the timeliness of imagery remains an issue. A 2013 study of 1k Google Street View panoramas and a 2019 analysis of 74k panoramas both found an average image age of two years~\cite{hara_tohme_2014, saha_project_2019}. While public infrastructure changes slowly, future systems must account for delays between image capture and access.

Another challenge is the ephemeral nature of elements such as cars, people, and temporary road closures visible in the imagery. With \system\, we used prompt engineering techniques~\cite{brown_language_2020} to mitigate this issue, but inaccuracies persisted. Future research should explore approaches to handling transient objects and preventing outdated information.

Finally, the vehicle-based point of view poses difficulties, as converting from a car to a pedestrian viewpoint adds cognitive load. Many participants emphasized the need for descriptions aligned with a pedestrian’s perspective to make street view imagery truly useful for navigation.

\subsubsection*{\textbf{Implications for MLLMs.}}
\system’s AI agent, powered by MLLMs, provides BLV users with access to street view imagery. Our user evaluations show promise for this approach, emphasizing the value of access to real-world information. Our technical evaluations highlight areas of accuracy, but also key limitations with MLLMs.

A major concern was the lack of spatial precision. Participants emphasized that merely knowing an object’s presence is insufficient; precise information about its location is crucial for navigation decisions. Our evaluations indicate that MLLMs fall short in reasoning about space with the precision required for blind navigation, where users rely entirely on environment descriptions. Additionally, MLLMs struggle to describe spatial relationships across multiple street view images, a vital component for facilitating real-world navigation. Future research should prioritize enhancing MLLMs’ spatial reasoning capabilities, especially in handling street view imagery, given its resemblance to real-world scenarios. Recent work has proposed benchmarks for evaluating visual-spatial intelligence~\cite{yang_thinking_2024, yang_v-irl_2024, liu_spatialcot_2025, chen_touchdown_2019, shiri_empirical_2024}.

While these models extracted useful information from street view imagery, they also sometimes made unwarranted assumptions about BLV users’ capabilities, leading to frustration among participants. This highlights the need for better understanding MLLMs' biases, as accessibility-related data is often underrepresented in training sets~\cite{bennett_its_2021, glazko_identifying_2024}. Future work should focus on adapting MLLMs for task-specific scenarios, ensuring that they can enable disability-aware interactions.

\section{Limitations}
\label{sec:limitations}
Our work explores how AI agents can make street view imagery accessible to BLV users and provides key insights into their needs and preferences. However, our study has several limitations.

First, our participants were employees of a major technology company, likely with some familiarity with MLLMs.
Their reactions may not reflect those of the broader BLV population, particularly those less familiar with AI advancements. Additionally, only one of our participants (P4) identified as low-vision, and therefore our user study does not reflect the full spectrum of visual ability. Future work should explore how a more diverse set of people with varying visual ability, technology familiarity, neurodiversity, and age range might perceive these systems. Second, due to time constraints, participants interacted with a limited set of examples. Long-term, real-world deployment is needed to understand how users adapt to the system’s limitations and discover new use cases. Additionally, our study focused on perceived usefulness rather than actual navigation outcomes, as participants did not physically navigate environments using the descriptions. Future research should assess how access to street view imagery impacts real-world navigation and explore in-situ, real-time street view integration. Last, our work explored two interaction modes: Route Preview and Virtual Exploration. Although these were most frequently mentioned in prior work and in feedback from BLV colleagues, other interaction methods may also be valuable. The scope of this work did not fully explore the design space for blind-accessible street view imagery and future research should develop a more comprehensive framework. Despite these limitations, our work serves as an initial step toward understanding how AI agents can provide BLV users access to street view imagery and identifies considerations for designing such systems.

\section{Conclusion}
\label{sec:conclusion}
We explored how AI-powered agents can make street view imagery accessible to BLV users, introducing opportunities for pre-travel navigation assistance. Our user study demonstrated that \system\ effectively surfaces meaningful environmental details, enabling BLV users to engage with street view imagery for both route previews and virtual exploration. Participants expressed strong interest in incorporating this information into their navigation workflows, but also highlighted challenges such as spatial imprecision and assumptions about user capabilities. Our technical evaluation further assessed the accuracy and reliability of MLLMs in extracting relevant information from street view imagery. We also discuss key insights from our findings and identify areas for improvement, including refining spatial reasoning, enhancing interaction design, and mitigating biases in AI-generated descriptions. While further research is needed, our work marks an initial step toward exploring how street view imagery can provide more information-rich and immersive navigation experiences for BLV users.


\begin{acks}
We thank our user study participants for their time and contributions. We greatly appreciate Amanda Swearngin, Griffin Dietz Smith, Jeffrey Bigham, Jessie Braden, Lilian de Greef, Ross Anderson, Samuel White, and Sarah Harling for their support and insightful feedback.
\end{acks}

\balance
\bibliographystyle{ACM-Reference-Format}
\bibliography{main}

\balance

\appendix
\section{Prompts}
\label{sec:prompts}

Here, we provide the main prompts used to facilitate \system's two interaction modes.

\subsection{Descriptions along the route}
\subsubsection{Prompt for describing non-intersection segments.}
{\small
\begin{verbatim}

[Role]
You are StreetDescriberGPT, an expert in giving description 
from old street view images for current navigation.

[Task Description]
Given the previous description, nearby points of interest, 
and three images for the left, front, and right side of a 
street, describe the sidewalk with information that would 
be helpful for a blind individual walking on the path. Be 
sure to include any changes in the street view images 
compared to the previous description. Note that the images 
are from a time when the images were taken and may not 
reflect the current state of the place. Thus, it is 
important to provide information that is likely to remain 
consistent over time. For example, when you see 
construction, it is possible that it may already have been 
completed. Always include information about specific 
places that you see, referring to their names from the 
board that appears in the images. Also include information 
about the street signs. Keep descriptions concise and 
relevant to walking navigation given the context. Do not 
repeat information from previous descriptions, but 
highlight the changes. For example, if the previous 
description says there is a mural on the left, do not 
mention the mural again. Always mention the nearby places, 
their direction, and distance in meters (do not use 
contractions). Always include accessibility-related 
information for sidewalks like width, changes in texture, 
obstacles, and mobility cues. Be as specific as you can 
be. Do not make any explicit mentions that the 
descriptions will help a blind or visually impaired 
person. You must respond in the following JSON format:
{{
"long_description":   <longer description: super detailed>,
"medium_description": <medium description: very concise 
                      but includes all the important 
                      information>,
"short_description": <short description: includes only the 
                     main information in one sentence>
}}

[Example]
For given set of images showing the left, front, and right 
view of the street, nearby points of interest, previous 
descriptions "..., sheltered bus stop on the left, street 
parking on both sides... the sidewalk is wide, ...", the 
output might include descriptions like: 

"A bookstore named Book Worm is on the left and the 
Concordia cafe is on the right sidewalk. The sidewalk 
appears to get narrower. The sheltered bus stop reads C 
line to 96th St." Respond in three different levels of 
verbosity - long, medium, and short. The long description 
should be detailed, medium is a concise version of the 
long description, and the short description only includes 
essential information in a sentence. The new description 
should be coherent with the previous descriptions. Always 
mention the nearby places, their direction, and distance. 
Always include signs and board names in the descriptions.

[Input]
Images: Images with a view of the street.
Previous Description: {prev_description}
Nearby Places: {nearby_places}

[Output]
Your description in the specified JSON format:
\end{verbatim}
}

\subsubsection{Prompt for describing intersections.}
{\small
\begin{verbatim}
[Role]
You are IntersectionDescriberGPT, an expert in giving 
description of an intersection from old street view images.

[Task Description]
Given images providing 360-degree view of that intersection
and nearby points of interest describe the intersection 
with information that would be helpful for blind 
individual crossing it. Note that the images are from a 
time when the images were taken and may not reflect the 
current state of the place. Thus, it is important to 
provide information that is likely to remain consistent 
over time. For example, when you see construction, it is 
possible that it may already have been completed. Always 
include information about specific places that you see, 
referring to their names from the board that appears in 
the images. Also include information about the street signs. 
Keep descriptions concise and relevant to walking 
navigation given the context. Always include accessibility-
related information about the presence or absence of 
audible pedestrian signals, tactile pavings, traffic 
lights and directions, sidewalk width, changes in texture, 
obstacles, and mobility cues. Be as specific as you can 
be. Never make any explicit mentions that it will help a 
blind or visually impaired person. Always mention the 
nearby places, their direction, and distance. You must 
respond in the following JSON format:
{{
"long_description": <longer description: super detailed>,
"medium_description": <medium description: very concise 
                      but includes all the important 
                      information>,
"short_description": <short description: includes only the
                     main information in one sentence>
}}

[Example]
For a given set of images showing 360-degree view of the
intersection and nearby points on interest, the output 
might include descriptions like:
"A four-way intersection with two-ways streets on both
sides and has a median. The intersection is controlled 
with accessible pedestrian signals. It is a busy 
intersection. Tactile pavings are present on all four 
sides of the intersection. Signs showing bus lanes and no 
parking can be seen as well." Respond in three different 
levels of verbosity - long, medium, and short. The long 
description should be detailed, medium is a concise 
version of the long description, and the short description 
only includes essential information in a sentence. The new 
description should be coherent with the previous 
descriptions. Always include signs and board names in the 
description. Always include accessibility-related 
information about the presence or absence of audible 
pedestrian signals and tactile pavings, sidewalk width, 
changes in texture, obstacles, and mobility cues. Always 
mention the nearby places, their direction, and distance 
in meters. Never make any explicit mentions that it will 
help a blind or visually impaired person.

[Input]
Images: Images with a view of the street.
Nearby Places: {nearby_places}

[Output]
Your description in the specified JSON format:
\end{verbatim}
}

\subsection{Descriptions near destinations}
{\small
\begin{verbatim}
[Role]
You are VisualPlaceDescriberGPT, an expert in describing
the visual elements of a place to a blind person that 
will help them navigate independently.

[Task Description]
Given the context of user's question, name of the place
they would like to go, and sequence of images on the 
path to the chosen place describe visual information 
that may help a blind person navigate to this place. 
Note that the images are from a time when the images 
were taken and may not reflect the current state of the 
place. Thus, it is important to provide information that 
is likely to remain consistent over time. For example, 
when you see construction, it is possible that it may 
already have been completed. Every detail you provide 
should be relevant to a blind person's navigation. 
Always include textual information from signboards and
accessibility-related information and spatial aspects of 
where the destination lies and what the entrance or 
structure looks like. Your response should be in the 
required JSON format:
{{
"path_summary": "Describe what the path looks like to
                 this place", 
"place_summary": "Describe the place with visual details
                  including the materials, colors, size, 
                  or anything that can help a blind
                  person navigating to this place.",
"mobility_cues": "Describe landmarks that appear on the
                  route to this place that may help a 
                  blind user who uses a white cane.",
"sidewalk": "Describe the sidewalk including its 
             material, width, changes in surface, or 
             anything noticeable that may be different
             from a usual sidewalk or help a blind 
             person navigate." 
"text": "Describe the text present on signages or boards
         near the place."
}}


[Example]
For the context "Is there a subway station here?", name
of place as "subway station entrance", and with sequence 
of images to this place, the output might be:
{{
"path_summary": "A curved path with wide sidewalk. Potted
                 plants on the right side and an open 
                 parking space on the left.",
"place_summary": "The subway station entrance has a set
                  of stairs going down with no elevators 
                  at this entrance. It is relatively 
                  narrow and in the middle of the 
                  sidewalk. There is a trash can right 
                  next to it on your way. The pillars 
                  are metallic and the stairs appear to 
                  be wooded.",
"mobility_cues": "There is a bicylce rack near the 
                  street on the sidewalk and then a 
                  trash can very close to the entrance.
                  Additionally, there are some potholes,
                  poles, and traffic signs if you pass 
                  the entrance. There seems to be a 
                  parking lot right before the subway 
                  station entrance.",
"sidewalk": "The sidewalk has concrete surface and is 
             medium sized, but gets wider as you 
             continue walking towards the entrance. It 
             also seems to curve a bit toward the street 
             as you keep walking. There are some bushes 
             close to the street on this sidewalk.",
"text": "The signage on the entrance reads: 1 train to 
         96th St, 2 min"
}}

[Input]
Context: {context}
Place: {place_name}

[Output]
Your response in the required JSON format:
\end{verbatim}
}

\subsection{Choosing directions for further exploration}
\subsubsection{Prompt for describing what lies in a direction.}
{\small
\begin{verbatim}
[Role]
You are PathDescriberGPT, an expert in describing what
lies in a specific direction based on a user-specified 
intention and old street view images.

[Task Description]
Given a set interntion, street name in a direction,
street's heading, the user's current heading, and the 
place that user is finding, describe what is or might be 
on a road that could help make the decision on whether to 
go in that direction or not. Include information both in 
support or against the possibility of finding the intented 
place. Keep the description concise and informative -- 1-3 
sentences only. Always start describing the road from the 
user's current heading on street name (\emph{e.g.} "Heading 
South on Adam Street: ...").

Your response should be in the following JSON format:
{{
"description": "Description of what is or is not in the
                direction (relevant to the place, both in 
                support and against)"
}}

[Example]
For the intention "find a grocery store" street name "Adam
St.", street heading "North", current heading "East", and 
and place type "grocery store", the output might be:
{{
"description": "Heading South on Adam Street: leads to a
                one-way residential street with houses and 
                trees. No commercial buildings or transit 
                stops in sight."
}}

[Input]
My Intention: {intention}
Street Name: {street_name}
Street heading: {new_heading}
Current heading: {curr_heading}
Place Type: {place_type}
Image: Refer to the given image.

[Output]
Your description in the required JSON format:
\end{verbatim}
}

\subsubsection{Prompt for suggestions on the most promising direction.}
{\small
\begin{verbatim}
[Role]
You are PathSelectorGPT, an expert in choosing the optimal
road from multiple candidates based on a user-specified 
intention and old street view images.

[Task Description]
Given a set intention, the road previously traveled, 
images of candidate roads and respective available 
candidate roads, select the best road from the crossroad. 
Always begin reason with heading and street name (\emph{e.g.} 
"Head South on Adam Street because..."). Your response 
should be in the following JSON format:
{{
"idx": "Selected road index (choose one from the range)", 
"reason": "Justification for your selection"
}}

[Example]
For the intention "find a grocery store", the road 
previously traveled as "1", and with candidates "2: Leads 
to residential area, 3: Leads to a shopping district", the 
output might be:
{{
"idx": "3", 
"reason": "Head South on Adam Street because a shopping
           mall is visible, making it more likely to have 
           a grocery store."
}}

[Input]
My Intention: {intention}
Road Descriptions: {road_descriptions}
Previously Traveled Road: Road {from_road_idx}
Images: Refer to given images.

[Output]
Your chosen road index and the reasoning behind your 
selection, in the required JSON format:
\end{verbatim}
}

\subsection{Descriptions along the block for exploration}
{
\small
\begin{verbatim}
[Role]
You are StreetDescriberGPT, an expert in giving description
of old street view images.

[Task Description]
Given a specified intention, primary destination, list of 
secondary cared labels, nearby points of interest, images 
providing a 180-degree view of the street, and all 
previous descriptions, describe the street view images 
with information that would be helpful for exploring this 
new space. Be sure to include any changes in the street 
view images compared to the previous description. Note 
that the images are from a time when they were taken and 
may not reflect the current state of the place. Thus, it 
is important to provide information that is likely to 
remain consistent over time. For example, when you see 
construction, it is possible that it may already have been 
completed. Always include information about specific 
places that you see, referring to their names from the 
board that appears in the images. Also include information 
about the street signs. Keep descriptions concise and 
relevant to walking navigation given the context. Do not 
repeat information from the previous description; 
highlight the changes. Always highlight information that 
may affect the accessibility of the sidewalks, such as 
width, changes in texture, obstacles, and mobility cues 
for blind people. Never make any explicit mentions that it 
will help a blind or visually impaired person.
{{
"long_description": <longer description: super detailed>,
"medium_description": <medium description: very concise 
                      but includes all the important 
                      information>,
"short_description": <short description: includes only the
                     main information in one sentence>
}}

[Example]
For intention, "reading a book," primary destination 
"bookstore," secondary labels "cafe, transit options, 
malls, crowds, parking", nearby points of interest, given 
set of images giving a 180-degree view of the street, 
previous descriptions "..., sheltered bus stop on the left 
sidewalk, closer to the street than to the building, 
street parking on both sides... the sidewalk is wide, 
...", the output might include descriptions like: 
"A bookstore named Book Worm is on the left sidewalk and 
the Concordia cafe is on the right sidewalk. The sidewalk 
appears to get narrower. The sheltered bus stop reads C 
line to 96th St." Respond in three different levels of 
verbosity - long, medium, and short. The long description 
should be detailed, medium is a concise version of the 
long description, and the short description only includes 
essential information in a sentence. The new description 
should be coherent with the previous descriptions. Always 
include signs and board names in the description. Always 
mention the nearby places, their direction, and distance.

[Input]
Intention: {intention}
Primary Place: {place_type}
Secondary Labels: {cared_secondary_categories}
Nearby Places: {nearby_places}
Previous Description: {prev_description}
Images: Images with a view of the street.

[Output]
Your description in the specified JSON format:
\end{verbatim}
}

\section{Route Preview Web Interface Example}
\label{sec:route_preview_ui_text}
Here, we include the text from Figure~\ref{fig:route_preview_ui}, which shows \system's web interface for the route preview interaction mode.

\subsection*{Descriptions en route to your destination along Westlake Avenue N going North}

\textbf{Instruction:} Review the descriptions as you move North on Westlake Avenue N. Please feel free to select a short, medium, or long version of the description.\\

\noindent\textit{At the Westlake Avenue N and Thomas Street intersection}.\\

\textbf{Short:} A four-way intersection with audible pedestrian signals, tactile pavings, wide sidewalks, and a nearby public transport stop 68 meters north. `California Closets' is 59 meters north.

\textbf{Medium:} A busy four-way intersection with two-way streets, tactile pavings, and curb ramps at all corners. Sidewalks are wide and unobstructed. Overhead electrical tram lines cross the intersection. Street signs indicate loading bus lanes and no parking zones.

\textbf{Long:} The intersection has high-rise office buildings with audible pedestrian signals at all corners. Tactile pavings assist navigation. Traffic lights are positioned diagonally. Sidewalks are wide and unobstructed. Overhead electrical tram lines are present. Curb ramps are available. `Westlake \& Thomas Stop' is 68 meters north, and `California Closets' is 59 meters left. Various street signs, a mural on a nearby building, and a for-lease sign are present.\\

\noindent\textit{43 meters further North, at 301 Westlake Avenue N}.\\

\textbf{Short:} On the left, a building with food illustrations and wide sidewalks. On the right, `Salt \& Pepper Deli Market' with parked cars. `Westlake \& Thomas Stop' is 42 meters left.

\textbf{Medium:} On the left, a building with colorful illustrations. A large tree provides shade. The sidewalk remains wide. On the right, a red brick building with `Salt \& Pepper Deli Market' has a large tree and a reserved parking sign.

\textbf{Long:} A glass-walled building with food illustrations is on the left. A large tree provides shade. On the right, `Salt \& Pepper Deli Market' has a tree in front and a reserved parking sign limiting access. The sidewalk is wide with no obstructions.

\subsection*{Detailed Visual Description of the Area Near Westlake \& Thomas Stop}

\textbf{Instruction:} Check out the detailed description of the path and sidewalk near Westlake \& Thomas Stop, including mobility cues and textual cues that appear in the street view imagery.

\subsubsection*{Description of Path Closer to Westlake \& Thomas Stop}
The path to the bus stop is straight along a bustling city street lined with modern buildings on both sides. The street has multiple lanes for cars and clearly marked crosswalks.

\subsubsection*{Visual Description of the Westlake \& Thomas Stop}
The bus stop is located along a wide sidewalk shaded by trees, with a glass canopy providing shelter. There is a metallic bench for seating and a sign displaying bus route and schedule information. The structure is primarily metallic with glass panels, featuring a `RapidRide' sign above the schedule post. A ticket vending machine is beside the bus shelter.

\subsubsection*{Mobility Cues}
Landmarks along the route include large office buildings with glass facades, intersection traffic lights, and clearly marked crosswalks. Trees are regularly spaced along the sidewalk, providing some shade.

\subsubsection*{Description of the Sidewalk Near Westlake \& Thomas Stop}
The sidewalk is made of concrete and is wide enough to easily navigate with a white cane. It appears to be in good condition, smooth, and without significant cracks or obstacles. There are marked pedestrian crosswalks with tactile paving at intersections.

\subsubsection*{Textual Cues Near Westlake \& Thomas Stop}
Signage near the bus stop includes the `RapidRide' sign above the scheduled post and street address numbers on buildings such as `320' near the entrance to the bus stop.

\section{Virtual Exploration Web Interface Example}
\label{sec:virtual_exploration_ui_text}

Here, we include the text from Figure~\ref{fig:virtual_exploration_ui}, which shows \system's web interface for the virtual exploration interaction mode.

\subsection*{Status}
\textbf{Ready!} Your intention is: I am buying a new house in this area and would like to see if this neighborhood is a quiet residential area with parks and good amenities. You are currently located at \textbf{Russell Street and Nassau Avenue intersection}. Go to the \textbf{"Description Keywords"} heading for next steps.

\subsection*{Description Keywords}
\textbf{Instruction:} Before you begin exploring, review the following keywords that will be highlighted in your descriptions. Please enter any additional keywords and hit the \textbf{Continue} button. You can add multiple keywords separated by commas.

\textbf{Following keywords will be highlighted in your descriptions of the street imagery as you explore:}

\begin{itemize}
    \item Parks
    \item Grocery stores
    \item Community centers
    \item Residential area
\end{itemize}

You can add multiple keywords separated by commas in the textbox below:  
\texttt{schools, public transport} (example input)  
\textbf{[Continue]}

\subsection*{Which Direction Would You Like to Explore Next?}
\textbf{Instruction:} Review the description for what lies ahead in each direction and the suggested direction. Hit the button to select the direction you'd like to explore next.

\textbf{Suggestion:} Head north on Russell Street because of the residential buildings, ample on-street parking, well-maintained sidewalks, and presence of trees, indicating a quieter residential area.

\subsubsection*{Exploration Options:}

\textbf{1. Heading North on Russell Street:}  
Residential buildings line both sides with ample on-street parking and well-maintained sidewalks. The presence of trees indicates some greenery, contributing to a quieter, residential feel. No visible commercial areas or parks in this view, suggesting a strictly residential zone with limited immediate amenities. \textbf{(Suggested direction)}

\textbf{2. Heading West on Norman Avenue:}  
The street features residential buildings and parked cars on both sides, indicating a residential area. The presence of both mid-rise apartment complexes and houses suggests a mixed residential environment. Limited green space in sight, so parks might be located further along or on adjacent streets.

\textbf{3. Heading East on Norman Avenue:}  
This area appears to be a mix of residential and commercial buildings with parked cars and some industrial presence. It might not be the quietest residential neighborhood, but it does have good amenities and street activity.

\section{Participant Demographics}
\label{sec:ptcpt_demographics}

Table~\ref{tab:main-study-ptcpts} summarizes our user study participants' demographics.

\begin{table*}[t]
\caption{Self-reported demographics of our user study participants. Gender information was collected as a free response. All participants, except P4, were totally blind. P4 reported having low vision due to severe central vision distortion. Participants rated their proficiency in using assistive technology (AT) for navigation on a scale of 1--5.}
\label{tab:main-study-ptcpts}
\centering
\renewcommand{\arraystretch}{1.2}
\begin{tabular}{@{}lccllll@{}}
\toprule
\textbf{PID} & \textbf{Age} & \textbf{Gender} & \textbf{Onset} & \textbf{Mobility Aids} & \textbf{Assistive Technology (AT) for Navigation} & \textbf{AT Proficiency} \\ \midrule

P1 & 50 & Male   & Age 40 & White cane & Apple/Google Maps, BlindSquare & (4/5) Very\\

P2 & 60 & Male   & At birth & White cane & Apple/Google Maps, BlindSquare, VoiceVista & (5/5) Extremely\\

P3 & 44 & Male   & Age 38 & White cane, Guide dog& Apple Maps, BlindSquare, Oko, GoodMaps  & (5/5) Extremely\\

P4 & 34 & Male & Age 17 & None & Apple/Google Maps, BlindSquare, Compass & (5/5) Extremely\\

P5 & 31 & Non-binary   & At birth & Guide dog& BlindSquare, VoiceVista, Oko  & (3/5) Moderately\\ 

P6 & 39 & Female   & Age 16 & Guide dog& Apple Maps, BlindSquare, GoodMaps  & (2/5) Slightly\\ 

P7 & 34& Female& Age 15& Guide dog& Apple Maps, BlindSquare, Soundscape& (3/5) Moderately\\ 

P8 & 31 & Female   & At birth & White cane & BlindSquare, Soundscape& (4/5) Very\\ 

P9 & 32 & Male   & At birth & Guide dog& BlindSquare, VoiceVista, Soundscape  & (5/5) Extremely\\ 

P10 & 51& Male& At birth & White cane, Guide dog& Apple/Google Maps, BlindSquare, Nearby Explorer& (5/5) Extremely\\ 

\bottomrule
\end{tabular}%
\end{table*}

\section{Questionnaires and Interview Questions}
\label{sec:study-ques}

\subsection{Pre-study Questionnaire}
\label{sec:pre-study-ques}

{\small 
\begin{itemize}
    \item[Q1.] What is your age?
    \item[Q2.] What is your gender?
    \item[Q3.] What is your occupation?
    \item[Q4.] Where do you live? (City, State)
    \item[Q5.] How often do you use screen readers (\emph{e.g.} VoiceOver) to access computers?
        \begin{itemize}
            \item[-] 1: Never
            \item[-] 2: Sometimes
            \item[-] 3: About half the time
            \item[-] 4: Most of the time
            \item[-] 5: Always
        \end{itemize}
    \item[Q6.] Which screen reader do you generally use for accessing computers?
    \item[Q7.] How would you rate your proficiency in using this screen reader?
        \begin{itemize}
            \item[-] 1: Not at all proficient
            \item[-] 2: Slightly proficient
            \item[-] 3: Moderately proficient
            \item[-] 4: Very proficient
            \item[-] 5: Extremely proficient
        \end{itemize}
    \item[Q8.] What is your degree of vision loss?
    \item[Q9.] Please describe your degree of vision loss.
    \item[Q10.] At what age did your vision impairments develop?
    \item[Q11.] Which mobility aids do you use?
    \item[Q12.] Which navigation apps do you typically use for outdoor navigation?
    \item[Q13.] How would you rate your proficiency in using assistive technology apps for outdoor navigation?
        \begin{itemize}
            \item[-] 1: Not at all proficient
            \item[-] 2: Slightly proficient
            \item[-] 3: Moderately proficient
            \item[-] 4: Very proficient
            \item[-] 5: Extremely proficient
        \end{itemize}
    \item[Q14.] How would you rate your confidence navigating outdoors without any sighted help (i.e., with assistive technology and mobility aids only)?
        \begin{itemize}
            \item[-] 1: Not at all confident
            \item[-] 2: Slightly proficient
            \item[-] 3: Moderately proficient
            \item[-] 4: Very proficient
            \item[-] 5: Extremely proficient
        \end{itemize}
    \item[Q15.] How would you rate your familiarity with street view imagery including how it is collected and what kind of information might be present in them?
        \begin{itemize}
            \item[-] 1: Not at all familiar
            \item[-] 2: Slightly familiar
            \item[-] 3: Moderately familiar
            \item[-] 4: Very familiar
            \item[-] 5: Extremely familiar
        \end{itemize}
    \item[Q16.] Do you currently use street view imagery to help with your outdoor navigation in any way? If so, how do you access it and for what reasons do you use it for?
\end{itemize}
}

\subsection{Post-prototype Questionnaire}
\label{sec:post-prototype-ques}

{\small
\begin{itemize}
    \item[Q1.] What did you like about this prototype?
    \item[Q2.] What did you dislike about this prototype?
    \item[Q3.] How useful did you find the information surfaced in descriptions by this prototype?
        \begin{itemize}
            \item[-] 1: Not at all useful
            \item[-] 2: Slightly useful
            \item[-] 3: Moderately useful
            \item[-] 4: Very useful
            \item[-] 5: Extremely useful
        \end{itemize}
    \item[Q4.] Elaborate on your rating. What specific types of information did you find useful and what additional pieces of information would you like the prototype to include?
    \item[Q5.] How relevant were the descriptions? Elaborate on your rating.
        \begin{itemize}
            \item[(a)] \textit{For the route-based prototype}: How well were the descriptions tailored to include accessibility-related information?
            \item[(b)] \textit{For the exploration-based prototype}: How well did the descriptions match your intention to explore the areas? Did it surface the information that you have liked to know given that scenario?
            \begin{itemize}
                \item[-] 1: Not at all relevant
                \item[-] 2: Slightly relevant
                \item[-] 3: Moderately relevant
                \item[-] 4: Very relevant
                \item[-] 5: Extremely relevant
            \end{itemize}
        \end{itemize}
    \item[Q6.] Since street view images can be old, how well do you think the prototype surfaced information that is likely to stay consistent? Elaborate on your rating.
        \begin{itemize}
            \item[-] 1: Not at all consistent
            \item[-] 2: Slightly consistent
            \item[-] 3: Moderately consistent
            \item[-] 4: Very consistent
            \item[-] 5: Extremely consistent
        \end{itemize}
    \item[Q7.] What level of detail in the descriptions between short, medium, long did you find the most useful? Rank them in order of usefulness.
    \item[Q8.] Explain your ranking. In which scenario do you imagine using each of these different levels of verbosity?
    \item[Q9.] How likely is it for you to use this prototype in environments that you're already familiar with?
        \begin{itemize}
            \item[-] 1: Not at all likely
            \item[-] 2: Slightly likely
            \item[-] 3: Moderately likely
            \item[-] 4: Very likely
            \item[-] 5: Extremely likely
        \end{itemize}
    \item[Q10.] How likely is it for you to use this prototype in environments that you're not at all familiar with?
        \begin{itemize}
            \item[-] 1: Not at all likely
            \item[-] 2: Slightly likely
            \item[-] 3: Moderately likely
            \item[-] 4: Very likely
            \item[-] 5: Extremely likely
        \end{itemize}
    \item[Q11.] Elaborate on the rating? How does familiarity affect the potential use of this prototype?
    \item[Q12.] How trustworthy would you say these descriptions are for you?
        \begin{itemize}
            \item[-] 1: Not at all trustworthy
            \item[-] 2: Slightly trustworthy
            \item[-] 3: Moderately trustworthy
            \item[-] 4: Very trustworthy
            \item[-] 5: Extremely trustworthy
        \end{itemize}
    \item[Q13.] How confident do you feel about knowing what to expect from the physical environment after using this prototype?
        \begin{itemize}
            \item[-] 1: Not at all confident
            \item[-] 2: Slightly confident
            \item[-] 3: Moderately confident
            \item[-] 4: Very confident
            \item[-] 5: Extremely confident
        \end{itemize}
    \item[Q14.] Elaborate on your rating.
    \item[Q15.] Do you have any suggestions for improving this prototype?
    \item[Q16.] If this prototype was made available to you right now, what would you use it for? Apart from the scenarios you experienced, what other scenarios can you imagine?
    \item[Q17.] Anything else you would like to add before we wrap up this prototype?
\end{itemize}
}

\subsection{Post-study Semi-structured Interview}
\label{sec:post-study-ques}
{\small
\begin{itemize}
    \item[Q1.] Now that you have tried both prototypes, how would you compare the two?
        \begin{itemize}
            \item[(a)] In which scenarios do you imagine using the route-based prototype?
            \item[(b)] In which scenarios do you imagine using the exploration-based prototype?
            \item[(c)] What do you think about the level of control that each agent offers? Do you have a preference?
        \end{itemize}
    \item[Q2.] Apart from these two prototypes, what new use cases can you imagine using street view imagery for?
        \begin{itemize}
            \item[(a)] What information from street view imagery would you like to access?
            \item[(b)] How would you like to access that information? In other words, what type of interaction/interface can you imagine while accessing that information?
        \end{itemize}
    \item[Q3.] Anything else you would like to add before we wrap up the study?
\end{itemize}
}

\section{Additional User Study Results}
\label{sec:additional_results}

\subsection{Ranking Preferences for Verbosity of Descriptions}
\label{sec:ranking_pref_appendix}

\begin{figure}[t]
    \centering
    \includegraphics[width=0.8\linewidth]{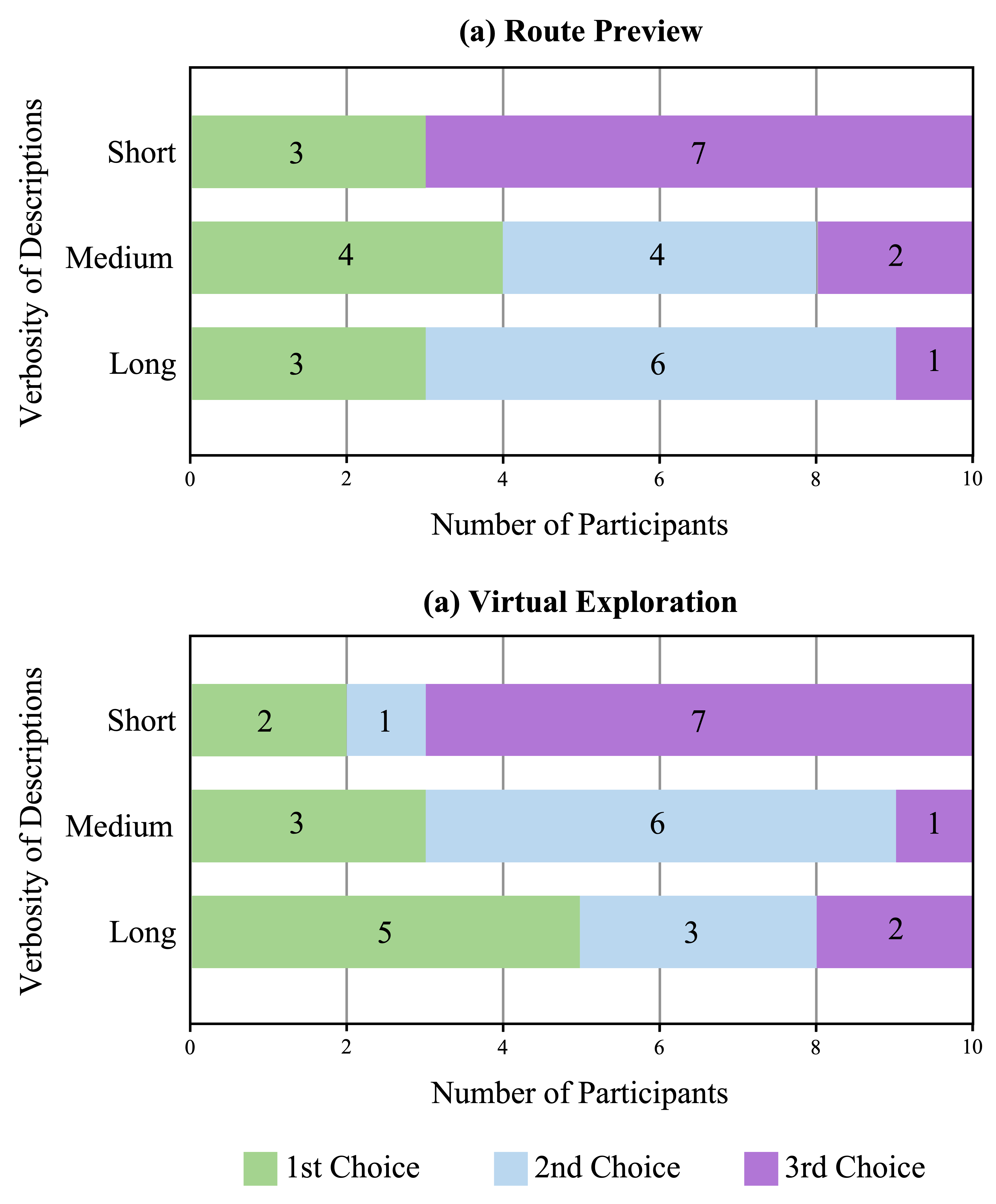}
    \caption{Participants’ ranking preferences ($N = 10$) for verbosity levels in (a) Route Preview and (b) Virtual Exploration. Preferences varied across participants: short and medium descriptions were more commonly preferred for Route Preview, while medium and long descriptions were favored for Virtual Exploration.}
    \label{fig:rankings_descriptions}
    \Description{Two bar charts reporting participants’ preferred verbosity levels—short, medium, and long—for Route Preview (top) and Virtual Exploration (bottom). Each bar is segmented by ranking: 1st choice (green), 2nd choice (blue), and 3rd choice (purple), with participant counts labeled inside each segment. For Route Preview, short and medium descriptions were most commonly ranked as 1st or 2nd choices. For Virtual Exploration, medium and long descriptions were more often preferred.}
\end{figure}

Figure~\ref{fig:rankings_descriptions} presents participants’ ranking preferences for verbosity levels across the two interaction modes. Short and medium descriptions were preferred for Route Previews, while medium and long descriptions were favored for Virtual Exploration. Participants valued the flexibility to adjust verbosity based on context, including time constraints, location familiarity, and visit purpose.

\subsection{Perceived Temporal Consistency of Descriptions}
\label{sec:results-temporal_consistency}

\begin{figure}[t]
    \centering
    \includegraphics[width=0.85\linewidth]{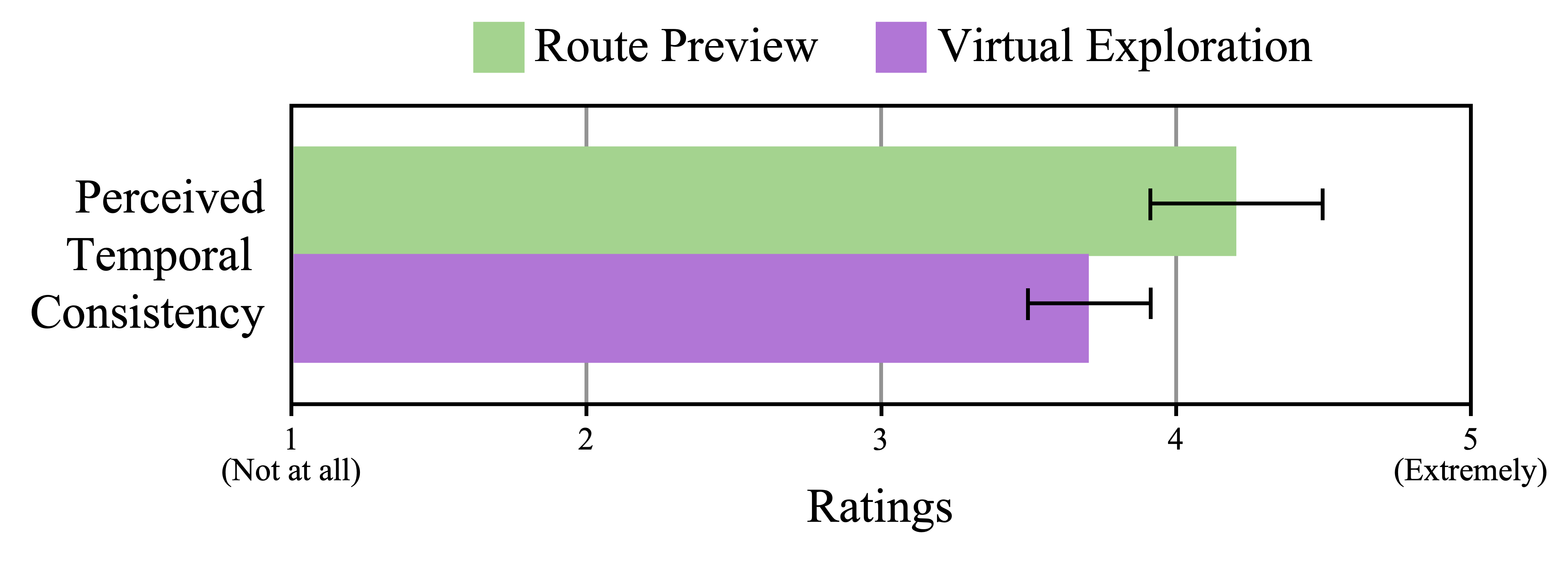}
    \caption{Participants’ average ratings ($N = 10$) of the perceived temporal consistency of descriptions across \system’s two interaction modes. While both received positive ratings, Route Preview was perceived to surface information more likely to remain consistent over time compared to Virtual Exploration. Error bars represent standard error.}
    \Description{Bar chart comparing average participant ratings for perceived temporal consistency of descriptions in Route Preview (green) and Virtual Exploration (purple). Route Preview received slightly higher ratings than Virtual Exploration. Ratings are plotted on a 1–5 scale, with exact mean values included in the main text.}
    \label{fig:ratings-temporal_consistency}
\end{figure}

Figure~\ref{fig:ratings-temporal_consistency} average ratings for participants' perceived temporal consistency of descriptions across the two prototypes. The mean ($\pm$ std. dev.) rating was $4.2$ ($\pm 0.9$) for route previews and $3.7$ ($\pm 0.7$) for virtual exploration. Participants noted  that some of the descriptions lacked temporal consistency. For example, the presence of transient objects like `Fedex' trucks and ongoing construction may be inaccurate when users eventually visit a location. While these inconsistencies sometimes caused confusion, participants recognized the inherent limitations of street view imagery. P6 noted: \textit{``I understand these are older images, so only major landmarks would be things I can 100\% trust, but there's no reason why I feel like the descriptions wouldn't be trustworthy.''} Despite these concerns, participants generally found the descriptions reliable for stable features like major landmarks and infrastructure.

\section{Annotation Instructions for Technical Evaluation}
\label{sec:annotation_instructions}

\begin{figure*}
    \centering
    \includegraphics[width=0.9\linewidth]{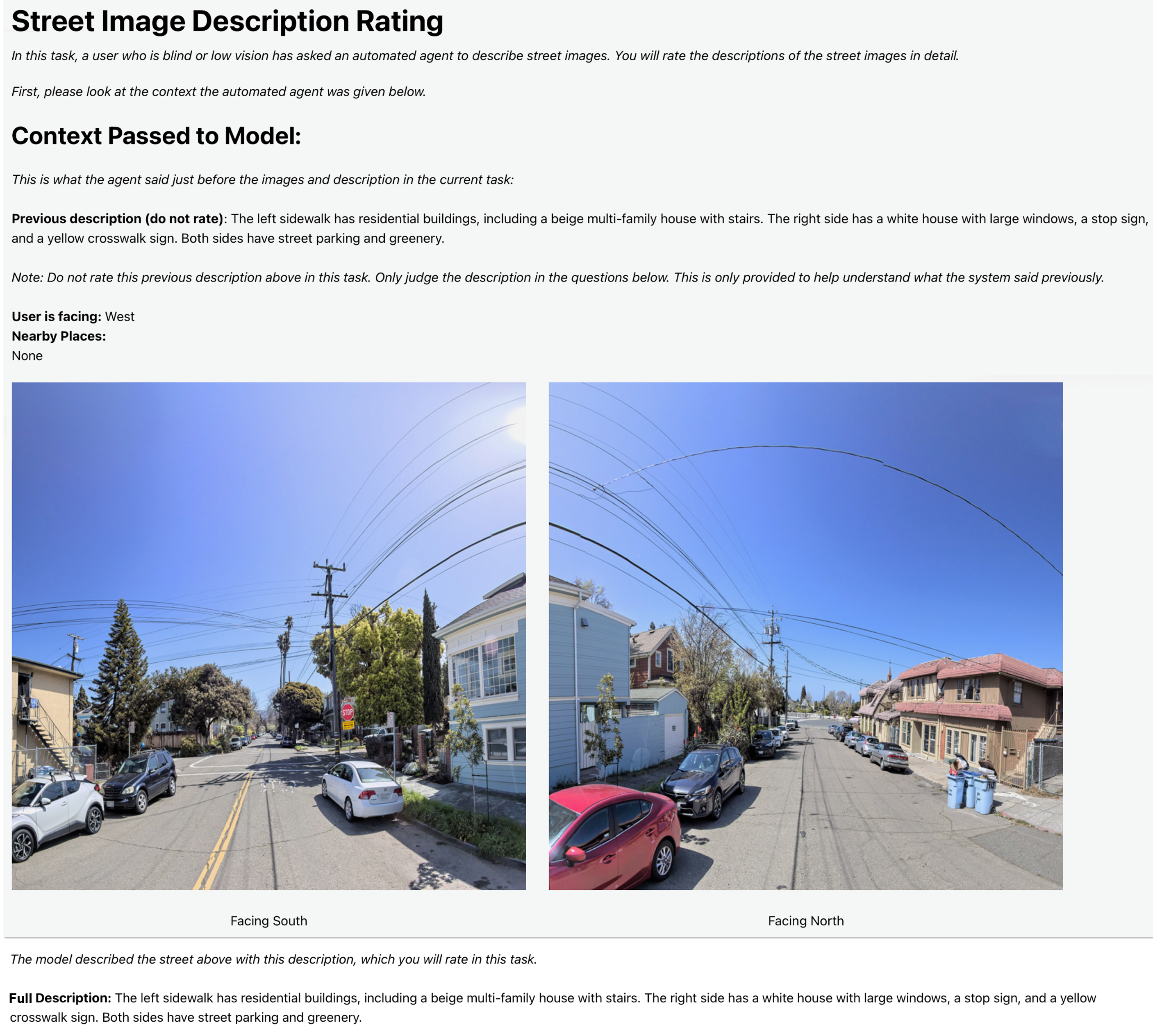}
    \caption{Annotation interface showing the context provided to the model before generating a description. This includes the previous sentence (not to be rated), user heading, nearby places, and corresponding street view imagery. Annotators use this context to evaluate each sentence in the full description.}
    \Description{Interface showing example street images and contextual information passed to the model before generating a description.}
    \label{fig:annotation_tag_task_context}
\end{figure*}


Figure~\ref{fig:annotation_tag_task_context} shows the context provided to annotators during the technical evaluation. This includes the previous description (for reference only), the user’s heading, nearby POIs, street view images, and the full system-generated description. Based on this context, annotators assess each sentence in the description for information type, correctness, consistency, and redundancy. The following evaluation form was used to rate each sentence according to these dimensions.

\subsection{Evaluation Form}

{\small
\noindent\textbf{Rate Each Sentence} \\
Now we will rate each sentence independently. Please read the sentence and answer each question based only on the information in that sentence. \\

\noindent\textbf{Sentence 1:} \textit{The left sidewalk has residential buildings, including a beige multi-family house with stairs}

\begin{itemize}
    \item [Q1.] \textbf{Type of information}: How subjective or objective is the description?
    \begin{itemize}
        \item [-] Subjective \\ (\emph{e.g.} busy street, well-maintained park)
        \item [-] Objective \\ (\emph{e.g.} signage reading 'No entry, parked cars)
        \item [-] Includes both subjective and objective information \\ (\emph{e.g.} a busy street with 'one-way' signage)
    \end{itemize}

    \item [Q2.] If description includes objective information, what type of objective information is this? (Select all that apply).
    \begin{itemize}
        \item [-] Points of Interest (POI) information \\ (a mention of nearby POl and possibly describing its relative location; \emph{e.g.} 'Shark Fitness' is in west, 72 meters away.)
        \item [-] Factual object information \\
        (a mention of objects visible in the images; \emph{e.g.} signage reading 'No entry, parked cars).
        \item [-] Accessibility-related information \\ (a mention of objects or scene description that is focused on providing more information about accessibility for blind or low vision individual; \emph{e.g.} sidewalk is wide and smooth, traffic signals are present, APS and curb cuts are visible)
        \item [-] Other (Free response)
    \end{itemize}

    \item [Q3.] \textbf{Accuracy}: How accurate is the description?
    \begin{itemize}
        \item [-] I cannot tell \\ (\emph{e.g.} states that a sign board reads "4th St," but the signage is partially occluded and thus its accuracy cannot not be verified)
        \item [-] Not correct \\ (\emph{e.g.} states that a bus stop is present, but it is neither visible in the images, nor is it mentioned in the context)
        \item [-] Partially correct \\ (\emph{e.g.} states that a bus stop is on the right; the bus stop is visible in the images, but context states it is on the left) 
        \item [-] Correct \\ (\emph{e.g.} states that a bus stop is present, which can is visible in the images or mentioned in the context information)
    \end{itemize}

    \item [Q4.] \textbf{Error type}: What type of error is in this sentence?
    \begin{itemize}
        \item [-] Plausible but not present visual detail \\ (\emph{e.g.} stating there is a factory near a parking lot of new cars, stating there is a small dog in a garden scene)
        \item [-] Plausible but inaccurate visual adjectives \\ (\emph{e.g.} describing a dark door as an open door, describing graffiti on building facade as a mural) 
        \item [-] Incorrect count / color / text or other factual information \\ (\emph{e.g.} a building features 3 doors when it actually features 4 doors, reading "Freeway" signage as "Figuoera Avenue")
        \item [-] Incorrect spatial detail \\ (\emph{e.g.} incorrectly stating a the trash can is behind the bus stop, when it is in front of it)
        \item [-] Complete hallucination \\ (\emph{e.g.} nothing in the image or context indicates even a remote possibility of this this statement being true)
        \item [-] Other (Free response)
    \end{itemize}

    \item [Q5.] \textbf{Consistency over time}: Given that the image was taken at least a couple months ago, how likely is it for the information provided in this image to stay consistent when a user visits the area shown in the image at a later time?
    \begin{itemize}
        \item [-] Not likely to remain consistent over time \\ (\emph{e.g.} a blue car is parked next to Starbucks entrance, a fedex truck is delivering packages)
        \item [-] Possibly remain consistent over time \\ (\emph{e.g.} busy street with cars parked along the street) 
        \item [-] Likely to remain consistent over time \\ (\emph{e.g.} bus stop has a shelter with glass panels, Starbucks is on the right, sidewalk is wide and smooth, curb cuts and APS are visible)
    \end{itemize}

    \item [Q6.] \textbf{Redundancy}: Given the previous description, does this sentence add any new information or repeat what has already been mentioned in the previous description?
    \begin{itemize}
        \item [-] No previous description provided \\ (Choose this if the previous description at the top of the task is blank)
        \item [-] Repeats information from previous description \\ (\emph{e.g.} both the previous description and this sentence state that the street is busy without adding any nuance or further information about the street)
        \item [-] Adds new information \\ (\emph{e.g.} states that a bus stop is now visible on the right, which wasn't mentioned in the previous description at all)
        \item [-] Updates previously mentioned information \\ (\emph{e.g.} states that a bus stop is now 10 meters away, which was 35 meters away as per the previous description)
    \end{itemize}
\end{itemize}
}


\end{document}